\documentclass[a4paper,11pt]{article}
\usepackage{jcappub} 
\usepackage{lineno}
\usepackage{orcidlink}
\usepackage{multirow}
\usepackage{subcaption}
\usepackage{hyperref}

\arxivnumber{2503.03823} 
\title{\boldmath Bounds on neutrino-DM interactions from TXS~0506+056 neutrino outburst}

 \author[1]{Gabriel~D.~Zapata\note{Corresponding author.}\orcidlink{0000-0002-2419-8424},}
 \author{Joel~Jones-P\'erez\orcidlink{0000-0002-2037-6369}}
 \author{and Alberto~M.~Gago\orcidlink{0000-0002-0019-9692}}
 \affiliation{Secci\'on F\'isica, Departamento de Ciencias, Pontificia Universidad Cat\'olica del Per\'u,\\Apartado 1761, Lima, Per\'u}

\emailAdd{gabriel.zapata@pucp.edu.pe}
\emailAdd{jones.j@pucp.edu.pe}
\emailAdd{agago@pucp.edu.pe}

\abstract{
We constrain the neutrino-dark matter cross-section using the $13 \pm 5$ neutrino event excess observed by IceCube in 2014-2015 from the direction of the blazar TXS~0506+056. 
Our analysis takes advantage of the dark matter overdensity spike surrounding the supermassive black hole at the center of the blazar. 
In our results, we take into account uncertainties related to the different types of neutrino emission models and the features of the dark matter spike, considering cross-sections that scale with energy as $\sigma \propto (E_{\nu}/E_0)^n$, for values of $n = 1,\, 0,\, -1,\, -2$.
In our best-case scenario, we obtain limits competitive with those derived from other active galaxies, tidal disruption events (TDEs), and the IC-170922A event.
}

\begin{document}

\maketitle

\flushbottom

\section{Introduction} \label{sec:intro}

The nature of dark matter (DM) has riddled the scientific community for more than half a century~\cite{Zwicky:1937zza, Rubin:1970zza, Planck:2018vyg, Cirelli:2024ssz}. In fact, despite the compelling gravitational evidence favoring the existence of DM, its mass, spin, and possible interactions with Standard Model (SM) particles remain unknown. Furthermore, no direct or indirect detection experiments~\cite{XENON:2018voc,LZ:2022ufs,SuperCDMS:2015eex,Lattaud:2022jnq,XENON:2023sxq,MAGIC:2016xys,HESS:2022ygk,Kawasaki:2021etm,Cui:2018klo} have provided any firm evidence of DM interactions with charged SM particles.

In front of this elusiveness of DM, one must also take into account what is probably the second most elusive particle we know of, that is, neutrinos. As is well known, for more than a decade neutrino oscillation experiments have indicated the need for non-zero neutrino masses~\cite{Super-Kamiokande:1998kpq, SNO:2011hxd, KamLAND:2004mhv,Gonzalez-Garcia:2007dlo}, contrary to the firm prediction of the SM. Thus, neutrinos are a sure window towards physics beyond the SM. In this sense, the possibility that the neutrino sector is the principal portal through which DM interacts becomes increasingly attractive~\cite{Arg_elles_2017, Blennow:2019fhy, Arguelles:2019ouk, Bell:2024uah}.

With this in mind, the observation of high-energy astrophysical neutrinos by detectors such as IceCube~\cite{Halzen:2022pez} presents a great opportunity to study these possible interactions. Neutrinos, unlike gamma rays and cosmic rays, interact only weakly with matter, allowing them to travel vast distances through space without being affected. However, if DM consists of particles that interact with neutrinos, the corresponding attenuation of astrophysical neutrinos could serve as a probe for these interactions.

The first object identified by IceCube as a high-energy neutrino source was blazar TXS~0506+056, following the observation of the IC-170922A neutrino event, which coincided with the direction of the blazar. Interestingly, observations by other experiments also indicated that TXS~0506+056 was experiencing a GeV gamma-ray flare~\cite{Aartsen_2018}, leading to a consistent description of the phenomena by blazar models~\cite{Cerruti:2018tmc, Liu_2019, Gasparyan_2021, Gao:2018mnu, Petropoulou:2019zqp}. The chance of coincidence of the neutrino with the flare of TXS~0506+056 is disfavored at the $3 \sigma$ level. In a posterior analysis, IceCube also found evidence of a neutrino outburst from the direction of TXS~0506+056 during the 5-month period between September 2014 and March 2015, with an excess of $13 \pm 5$ high-energy muon neutrino events respect to atmospheric backgrounds~\cite{Aartsen_2018b}. However, this neutrino emission was not accompanied by an electromagnetic flare which, as we shall see, provides a challenge to blazar models.

Other associations between high-energy neutrinos and astrophysical sources have since been made, including the active galaxy nuclei (AGN) NGC  1068~\cite{IceCube:2022der}, the blazar PKS 1741-038~\cite{Plavin:2022oyy}, and tidal disrupted events (TDEs)~\cite{Stein:2020xhk, Reusch:2021ztx, vanVelzen:2021zsm}.

All of this data compels an evaluation of neutrino-DM interactions. The possibility of use the neutrino event IC-170922A to put bounds on the neutrino-DM cross-section was first introduced in~\cite{Choi_2019}, where the authors considered the path of neutrinos through both the cosmological DM background and the Milky Way’s DM halo.\footnote{See also~\cite{Koren:2019wwi} for an early constraint based on neutrino and photon arrival times, and \cite{Fujiwara:2024qos} for a more recent study in this direction.}
Stronger constraints were later obtained by accounting for the dense DM spike in the center of TXS~0506+056, increasing the DM column density along the neutrino path by several orders of magnitude~\cite{Cline_2022,Ferrer_2023}. Additional limits on neutrino-DM interactions have been derived from other sources of high-energy neutrinos, such as the AGN NGC 1068~\cite{Cline_2023} and TDEs~\cite{Fujiwara:2023lsv}.
Bounds from such astrophysical neutrino sources have also been interpreted in specific models, see~\cite{Reynoso:2022vrn,KA:2023dyz} for examples.

In this study we focus on the blazar TXS~0506+056, but instead of the single IC-170922A event, we consider the $13 \pm 5$ high-energy neutrino events from the 2014–2015 neutrino outburst. As mentioned earlier, this phenomenon has been difficult to reproduce in blazar models, so it is interesting to consider what information we can get from those few models that do so, taking also into account the uncertainties related to the features of the DM spike. In addition, previous studies of other astrophysical high-energy neutrino events have considered simplified setups for neutrino-DM interactions, assuming constant and linearly energy-dependent cross-sections. In this work we also contemplate situations where the cross-section is inversely proportional to the energy and to the square of the energy, as motivated by models involving light scalar mediators.

This paper is organized as follows: In Section~\ref{sec:nu_flux_IC}, we describe the 2014-2015 neutrino outburst from TXS 0506+056 and the neutrino fluxes consistent with that observation. In Section~\ref{sec:dm_spike}, we analyze the blazar's DM density profile, emphasizing the properties of the DM spike and its effect on the column density. In Section~\ref{sec:nudm}, we incorporate the fluxes from Section~\ref{sec:nu_flux_IC} into the cascade equation to constrain the neutrino-DM cross-section under different energy-dependent cross-sections, accounting for uncertainties in the DM spike parameters. Here we also compare our constraints with existing limits in the literature. Finally, in Section~\ref{sec:concl} we present our conclusions.

\section{Neutrino flux from TXS~0506+056} \label{sec:nu_flux_IC}

Any bound placed on the neutrino - DM cross-section $\sigma_{\nu DM}$ ultimately depends on the neutrino flux emitted by TXS~0506+056 during the 2014-2015 neutrino outburst~\cite{Cline_2023, Cline_2022}. This means that the constraints will be inevitably tied to the specific blazar model for the flux. These models are usually either hadronic, leptonic or hybrid leptohadronic, depending on the dominant processes with the blazar jet. In our case, the observation of high-energy neutrinos points toward either a hadronic or leptohadronic description of the blazar, since leptonic models intrinsically produce too few neutrinos to be consistent with the outburst. However, in hadronic models the neutrino flux is generated via the decay of charged pions, while gamma rays are produced through the decay of neutral pions, resulting in similar numbers of neutrinos and gammas. This represents a challenge in explaining the 2014-2015 neutrino outburst, as it occurred during a period of low gamma ray emission. Thus, for hadronic models it is very difficult to generate a sufficiently large enough neutrino flux explaining the outburst without saturating the bound on the electromagnetic emission of the blazar. Unfortunately, leptohadronic models suffer from similar difficulties~\cite{Reimer:2018vvw,Rodrigues_2019,Petropoulou:2019zqp,Gasparyan_2021}.

\begin{figure} 
\centering
\includegraphics[width=0.49\textwidth]{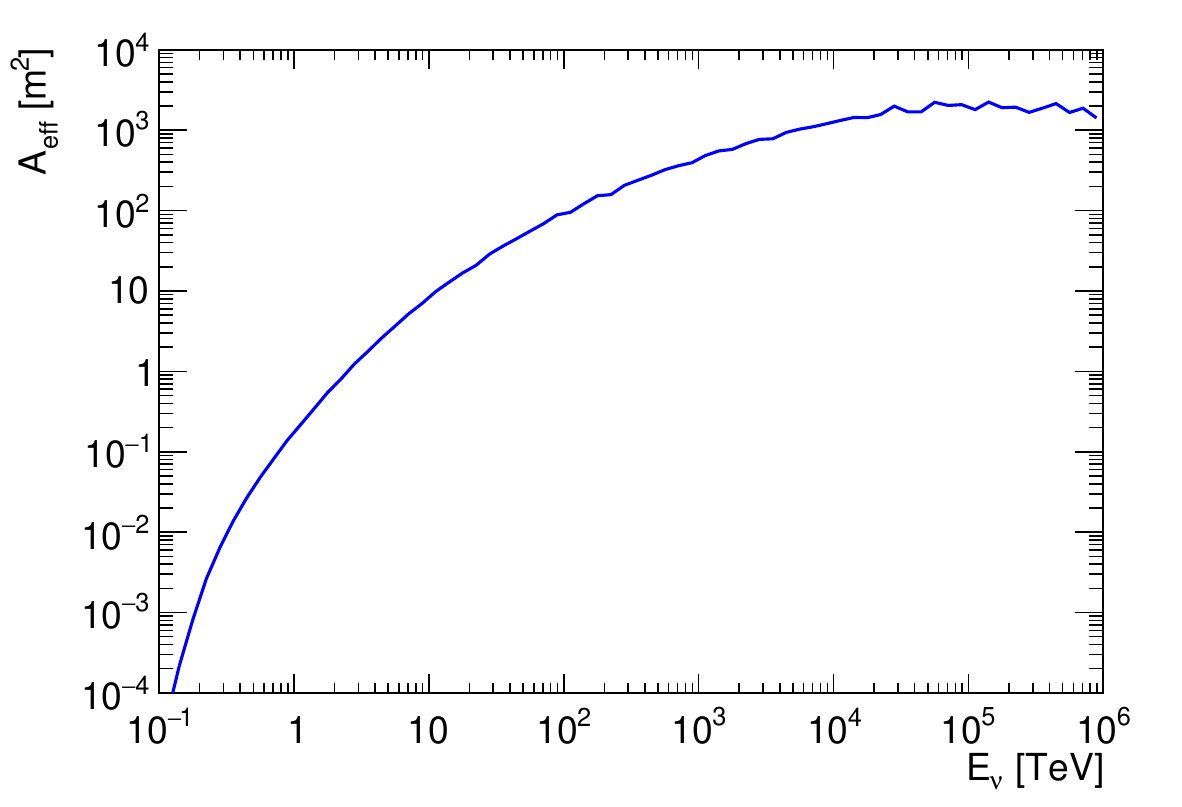}
\includegraphics[width=0.49\textwidth]{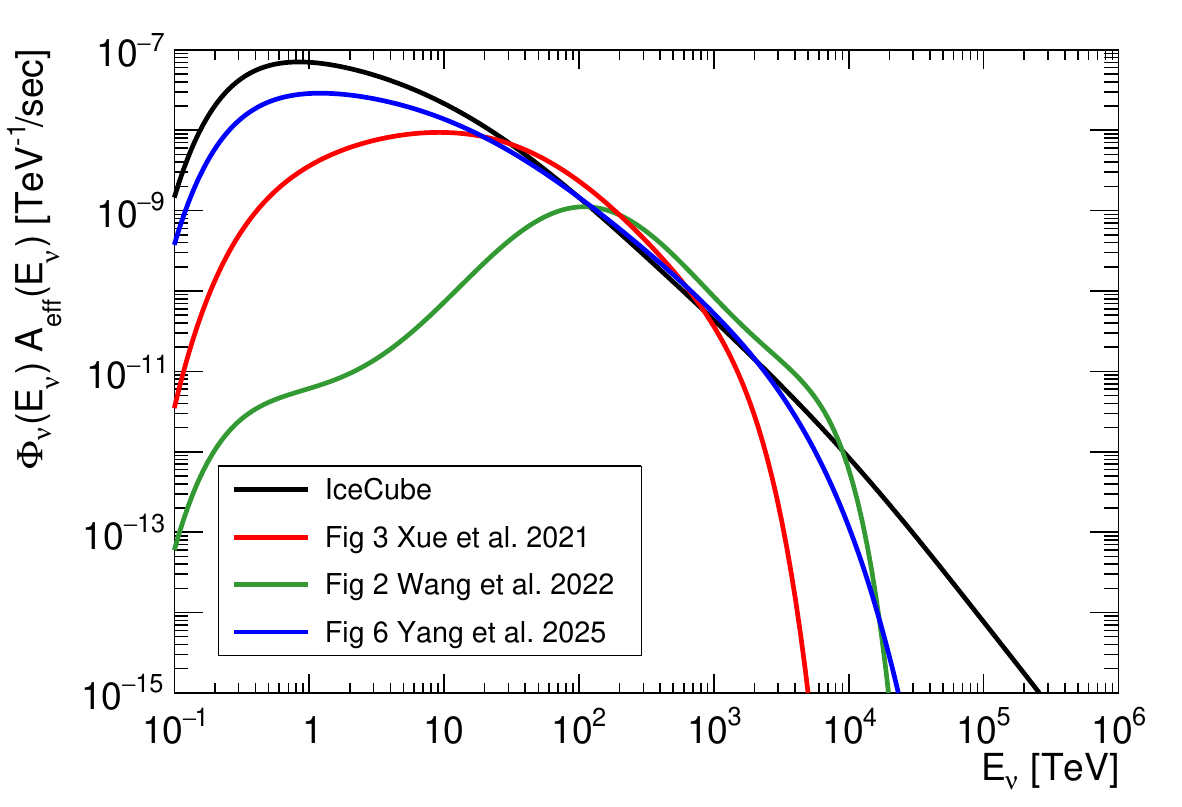}
\caption{\label{fig:aeff} Left: Effective area $A_{\textrm{eff}}$ from TXS~0506+056 in the sample IC86b, provided by IceCube~\cite{IceCube_2018}. Right: Spectrum of events to be expected at IceCube, for each of the different fluxes considered.}
\end{figure}

In the following, we will select models that address and overcome this incompatibility. As a first step, we need to assess which spectra are capable of describing the $13\pm5$ events at IceCube in the absence of neutrino-DM interactions. To this end, given a flux, the total number of muon neutrino events to be expected at IceCube can be calculated using:
\begin{equation} \label{eq:nevents}
N_{\textrm{pred}} = t_{\textrm{obs}} \int d E_{\nu}\, \Phi_{\nu} (E_{\nu})\, A_{\textrm{eff}} (E_{\nu})
\end{equation}
with $\Phi_{\nu} \equiv \Phi_{ \nu_{\mu} + \bar{\nu}_{\mu} }$ the neutrino flux arriving to the detector, and $t_{\textrm{obs}}$ the observation time. Furthermore, the effective detection area $A_{\textrm{eff}}$ encapsulates the probability of a neutrino generating a muon within the detector via weak interactions, depending on the neutrino energy, the detector geometry, and source direction. Following~\cite{Aartsen_2018b}, the 2014 - 2015 neutrino outburst occurred in the sample IC86b, with the effective area shown on the left panel of Figure~\ref{fig:aeff}~\cite{IceCube_2018}.

With Eq.~(\ref{eq:nevents}), we can perform a consistency check following the conclusions of~\cite{Aartsen_2018b}, where the collaboration reports that the $13\pm5$ events can be fit into an unbroken power-law of the form:
\begin{equation} \label{eq:flux_ic1}
\Phi_{ \nu_{\mu} + \bar{\nu}_{\mu} } (E_{\nu}) = \Phi_{\textrm{ref}} \left( \frac{E_{\nu}}{100 ~ \textrm{TeV}} \right)^{-\gamma},
\end{equation}
where the best-fit parameters have $\gamma=2.2 \pm 0.2$ and $\Phi_{\textrm{ref}} = 1.6^{+0.7}_{-0.6} \times 10^{-15}$ TeV$^{-1}$ cm$^{-2}$ sec$^{-1}$. These values are obtained with a time-dependent analysis with a box-shaped time window of duration $t_{\textrm{obs}}=158$~days, using an unbinned maximum likelihood ratio method to search for an excess on the number of neutrino events consistent with a point source, with $\gamma$, $\Phi_{\textrm{ref}}$ and $t_{\textrm{obs}}$ being fitting parameters. 
Replacing the parameters above in Eq.~(\ref{eq:nevents}), we obtain $N_{\textrm{pred}} \approx 15$, integrating in the energy range $E_{\nu} = 10^{-1} - 10^6$ TeV.
Our predicted number of events, with its uncertainties, is consistent with the $13 \pm 5$ neutrinos observed by IceCube, validating our procedure.

In the literature, there exist several models claiming to be consistent with the neutrino outburst~\cite{Xue_2019,Zhang_2020,Xue_2021, Wang_2022,Yang:2024bsf, wang2024unified,KhateeZathul:2024tgu}, managing to generate a large enough neutrino flux without exceeding the observed gamma ray emission. In the following, in order to place the most conservative bounds on $\sigma_{\nu DM}$, we focus on the three models giving the largest $N_{\textrm{pred}}$ for $\sigma_{\nu DM}=0$, as calculated with Eq.~(\ref{eq:nevents}). The models under consideration all have $N_{\textrm{pred}}>6.55$, which is the 90\% C.L. lower limit on the number $13 \pm 5$ of observed IceCube events, assuming a normal distribution.

The first model is the two-zone radiation model, also called the inner-outer blob model, of Xue~\textit{et al.}~\cite{Xue_2019,Xue_2021}. In this model, neutrino and gamma-ray emission occur in a first zone (inner blob) close to the Schwarzschild radius of the black hole, where X-ray photons from the hot corona absorb all gamma rays. The second zone (outer blob), located farther away, is responsible for the less energetic multi-wavelength electromagnetic emission observed. This flux is shown in Figure~3 of~\cite{Xue_2021} and has $N_{\textrm{pred}}=11.5$.

The second model, from Wang~\textit{et al.}~\cite{Wang_2022} considers the possibility of having the neutrinos being generated by the interactions between the jet of the blazar and a dense cloud, the latter originating from the envelope of a Red Giant star being tidally disrupted by the black hole. In this case, interactions of low-energy protons provide an electromagnetic spectrum that is spread out, and thus help respect the bounds, while the high energy undeflected protons lead to the neutrino flux. Case 2, shown in Figure~2 of the paper, gives $N_{\textrm{pred}}=6.83$.

The third model we consider is a recent work proposed by Yang~\textit{et al.}~\cite{Yang:2024bsf}. Here, similarly to~\cite{KhateeZathul:2024tgu}, the neutrino emission is attributed to the accretion flow of the SMBH rather than the relativistic jet. To account for the 2014–2015 neutrino outburst, a super-Eddington accretion rate is required. The authors examine neutrino production in both the magnetically arrested disk (MAD) and the standard and normal evolution (SANE) accretion regimes. While both scenarios yield a neutrino flux consistent with observations, we adopt the flux from the SANE regime, as it is associated with less energetic jets, making it more compatible with the low gamma-ray emission observed during the 2014–2015 neutrino outburst.\footnote{In the recent work \mbox{\cite{Fiorillo:2025cgm}}, the authors point out that the X-ray and proton luminosities assumed in these models are significantly higher than expected, and when realistic parameters are used, the resulting coronal neutrino emission from TXS 0506+056 is too low to account for the IceCube observation.} The SANE case, shown in the Figure~6 of~\cite{Yang:2024bsf} with particle acceleration efficiency $\eta = 300$, gives $N_{\textrm{pred}}=13.3$.

The spectra of events to be expected from each flux at IceCube is shown on the right panel of Figure~\ref{fig:aeff}. This corresponds to the convolution of the flux and effective area, namely, the integrand of Eq.~(\ref{eq:nevents}). These are compared with the IceCube flux, Eq.~(\ref{eq:flux_ic1}), where it is clear that the latter has a larger contribution from low-energy events than any of the models considered here, as well as a long tail. We find that the flux from Xue~\textit{et al.}~\cite{Xue_2021} has a large contribution from medium energy neutrinos, peaking around $E_\nu\sim10$~TeV, but being around two orders of magnitude smaller than IceCube at low energy, and very strongly suppressed for energies above $\sim10^3$~TeV. In contrast, the flux from Wang~\textit{et al.}~\cite{Wang_2022} leads to a much harder neutrino spectrum, peaking around $E_\nu\sim100$~TeV, and a tail going beyond that from Xue~\textit{et al.}~\cite{Xue_2021}. Nevertheless, the noticeable lack of low-energy neutrinos leads to a relatively small $N_{\textrm{pred}}$. Finally, The flux from Yang~\textit{et al.}~\cite{Yang:2024bsf} has a contribution from the low-energy neutrinos similar to the IceCube flux, peaking around $E_{\nu} \sim 1$ TeV, but with a slightly shorter tail.
It is worth noting that although the neutrino spectra have different shapes, given the significant fractional uncertainties in the energy reconstruction of the observed neutrinos (${\cal{O}}(25 \%)$ in $\log_{10}(E_{\nu}/\mbox{TeV})$),\footnote{Derived from the plot S6 in the Ref. \cite{Aartsen_2018b}.} it is not possible to use the energy dependence of the IceCube flux to discriminate any of the considered models.

\section{DM spike} \label{sec:dm_spike}

The second key element needed to constrain $\sigma_{\nu DM}$ is the DM density profile of TXS~0506+056 and its host galaxy.
If the growth of the black hole is adiabatic, an initially cuspy dark matter profile of the form $\rho (r) = \rho_0 (r/r_0)^{-\gamma}$, with $0<\gamma<2$ and $\rho_0$ a reference density at $r=r_0$, evolves into a steeper distribution~\cite{Gondolo_1999, Ullio:2001fb}:
\begin{equation}
\label{eq:rhoprime}
\rho'(r) = \rho_R\, g_{\gamma}(r) \left( \frac{R_{\textrm{sp}}}{r} \right)^{\gamma_{\textrm{sp}}}~, \qquad (4R_S < r \leq R_{\textrm{sp}})~.
\end{equation}
Here, we have $\gamma_{\textrm{sp}}=(9-2 \gamma)/(4-\gamma)$ as the spike slope, which is valid up to the spike radius $R_{\textrm{sp}}=\alpha_{\gamma}\,r_0(M_{BH}/\rho_0 \,r_0^3)^{1/(3-\gamma)}$. Apart from the black hole mass $M_{BH}$, the spike radius depends on a normalization factor $\alpha_\gamma$, which must be obtained numerically (values are given in~\cite{Gondolo_1999}). In addition, $\rho'(r)$ depends on an additional function $g_\gamma(r)$, which is also obtained numerically in~\cite{Gondolo_1999}, and a further normalization factor $\rho_R = \rho_0 (R_{\textrm{sp}}/r_0)^{-\gamma}$ designed such that $\rho'(R_{\textrm{sp}})=\rho(R_{\textrm{sp}})$. Finally, Eq.~(\ref{eq:rhoprime}) is valid only if $r\geq 4R_S$, where $R_S$ is the Schwarzschild radius, with particles in smaller orbits being accreted into the SMBH, leading to a vanishing distribution. 

For definiteness, in the following we take $\gamma = 1 $, which implies $ \gamma_{\textrm{sp}} = 7/3$, $\alpha_{\gamma}=0.122$ and $g_{\gamma}(r) \approx (1- 4R_S/r)^3$.
This election for the slope $\gamma = 1$ corresponds to an initial Navarro-Frenk-White (NFW) density profile~\mbox{\cite{Navarro:1995iw, Navarro:1996gj}}, given by
\begin{equation}
\label{eq:rho}
\rho_{NFW}(r) = \rho_0  \frac{r_0}{r}  \left( 1+\frac{r}{r_0} \right)^{-2}.~
\end{equation}
We consider that outside the spike radius $R_{\textrm{sp}}$, the density of dark matter halo is still determined by the pre-existing NFW profile.

For a fixed $r_0$, $\gamma_{\textrm{sp}}$ and $M_{BH}$, the reference density $\rho_0$ determines the spike size $R_{\textrm{sp}}$. The density can be estimated by requiring that the enclosed dark matter mass be roughly equal to the black hole mass~\cite{Ullio:2001fb, Wang:2021jic, Granelli_2022ysi, Bhowmick:2022zkj}
\begin{equation} \label{eq:norm_rho1}
\int_{r_{\textrm{min}}}^{r_{\textrm{max}}} 4 \pi\, \rho'(r)\, r^2 dr \approx M_{BH}
\end{equation}
with $r_{\textrm{min}} = 4 R_S$ and $r_{\textrm{max}} = 10^5 \,R_S$, the radius of influence of the SMBH. Then
\begin{equation}
\rho_0 = \left( \frac{(3-\gamma_{\textrm{sp}})  M_{BH}}{4 \pi \,r_0^{\gamma}\, \xi^{(\gamma_{\textrm{sp}}-\gamma)}\, \left(r_{\textrm{max}}^{3-\gamma_{\textrm{sp}}}-r_{\textrm{min}}^{3-\gamma_{\textrm{sp}}}\right)} \right)^{(4-\gamma)}
\end{equation}
with $\xi = \alpha_{\gamma}\,r_0\,(M_{BH}/ r_0^3)^{1/(3-\gamma)}$. 
We adopt $r_0 = 10$ kpc as the scale radius~\cite{Ferrer_2023, Fujiwara:2023lsv}, comparable to that of the Milky Way, for which $r_0 \sim 20$ kpc~\cite{Cautun:2019eaf}. The mass of the black hole at the center of TXS~0506+056 was estimated in~\cite{Padovani_2019} to be $M_{BH} \approx 3 \times 10^8 M_{\odot}$, leading to $\rho_0 \approx 2 \times 10^5$ GeV/cm$^3$. This, in turn, implies that $R_{\textrm{sp}}\approx0.3$~pc.

A more precise description of the DM spike would require the inclusion of relativistic effects. In \mbox{\cite{Sadeghian:2013laa}} it was found that the relativistic effects reduce the inner radius of the spike from $4 R_S$ to $2 R_S$, and increase significantly the density profile near the black hole. This enhancement of the density is larger for the case of a rotating black hole \mbox{\cite{Ferrer:2017xwm}}.
However, these relativistic effects are relevant only close to the SMBH, at $r \lesssim 20 R_S$, while, as we will see, neutrinos are produced in more distant areas, so we will disregard such effects.

Having defined the dark matter density profile for all values of $r$, we now consider the effects of dark matter self-annihilation. Such a possibility is to be expected once one allows $\nu\, DM\to \nu\, DM$ scattering, however, in the following we do not relate $\sigma_{\nu DM}$ with the annihilation rate. In this case, the profile is suppressed by a factor $\rho_{\textrm{sat}} = m_{DM}/( \langle \sigma v \rangle_{\textrm{ann}}\, t_{BH})$, where $m_{DM}$ is the DM mass, $\langle \sigma v \rangle_{\textrm{ann}}$ is its velocity averaged annihilation cross section and $t_{BH}$ is the age of the black hole, for which we take the value $t_{BH} = 10^9$ yr. In this situation, one needs to replace: 
\begin{equation}
    \rho_{DM}(r)\to\frac{\rho_{DM}(r)\,\rho_{\textrm{sat}}}{\rho_{DM}(r)+\rho_{\textrm{sat}}}
\end{equation}
where $\rho_{DM}(r)$ represents either $\rho'(r)$ or $\rho_{NFW}(r)$, depending on $r$.\footnote{Some works have pointed out that, considering general particle velocity distributions, the DM density profile cannot shallower than $\sim r^{-1/2}$ \mbox{\cite{Vasiliev:2007vh, Shelton:2015aqa, Shapiro:2016ypb}}.
To simplify the discussion, we take the more cored DM profile as a conservative choice.} It is important to note that the original profile is recovered when $\rho_{\textrm{sat}}\to\infty$, which happens for vanishing $\langle\sigma v\rangle_{\textrm{ann}}$ or very large $m_{DM}$.
To explore the range of possible outcomes for different annihilation cross-sections, we follow~\cite{Wang:2021jic, Granelli_2022ysi} by considering three benchmark models BM1 - BM3, with $\langle \sigma v \rangle_{\textrm{ann}} = (0,\, 0.01,\, 3) \times 10^{-26}$~cm$^3/$s, respectively.
As noted in \mbox{\cite{Granelli_2022ysi}}, the first benchmark with $\langle \sigma v \rangle_{\textrm{ann}}=0$ may be appropriate for asymmetric dark matter models \mbox{\cite{Petraki:2013wwa, Zurek:2013wia}} and, more generally, for scenarios where no significant spike depletion is expected. 
The third benchmark with $\langle \sigma v \rangle_{\textrm{ann}}=3 \times 10^{-26}$~cm$^3/$s corresponds to thermal relic dark matter, while the second with $\langle \sigma v \rangle_{\textrm{ann}}=10^{-28}$~cm$^3/$s represents an intermediate annihilation rate.

In addition to annihilation effects, gravitational interactions between DM and stars surrounding the black hole may deplete the structure of the spike. Depending on the age of the galactic bulge, the spike can relax to a profile with an index as low as $\gamma_{\textrm{sp}} = 3/2$, without modifying aforementioned parameters, such as $\alpha_\gamma$ and $\rho_0$~\cite{Gnedin:2003rj}.
To be conservative, we will consider the models BM1$'$ - BM3$'$ using this less cuspy value $\gamma_{\textrm{sp}} = 3/2$, as was also done in~\cite{Cline_2022}.

\begin{figure}
\centering
\includegraphics[width=0.49\textwidth]{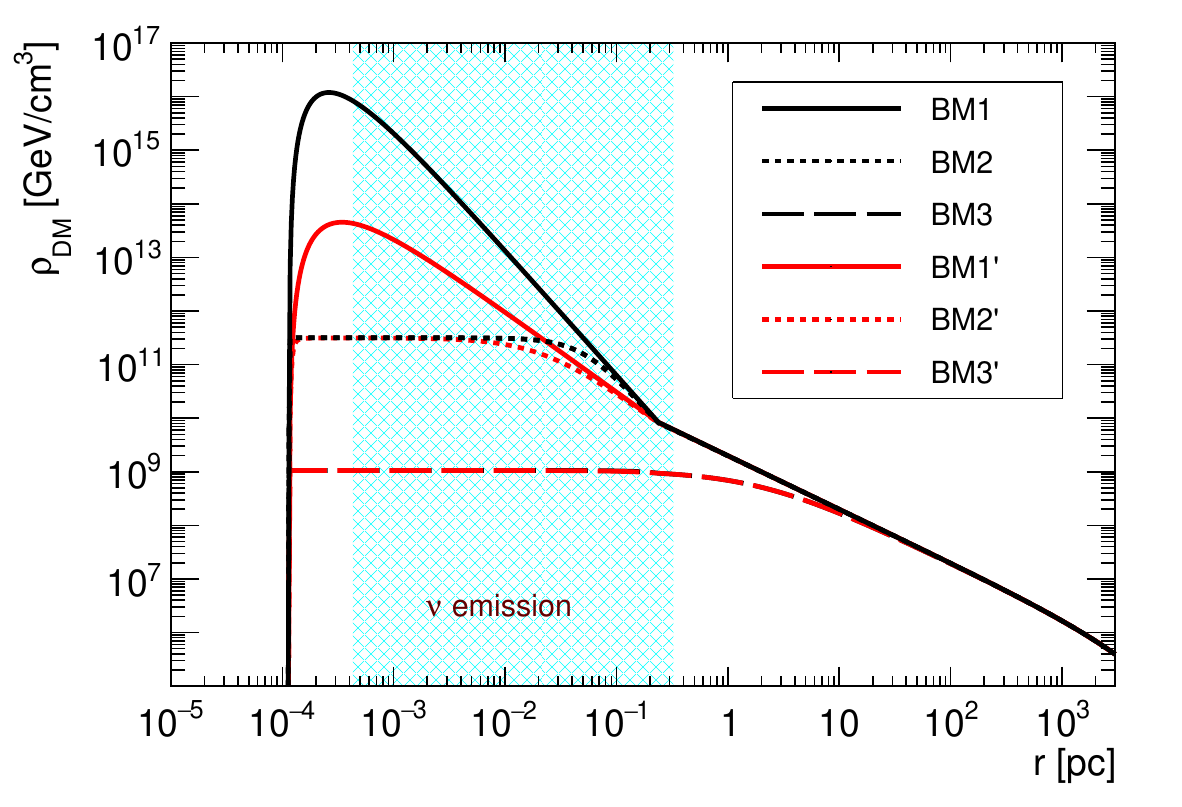}
\includegraphics[width=0.49\textwidth]{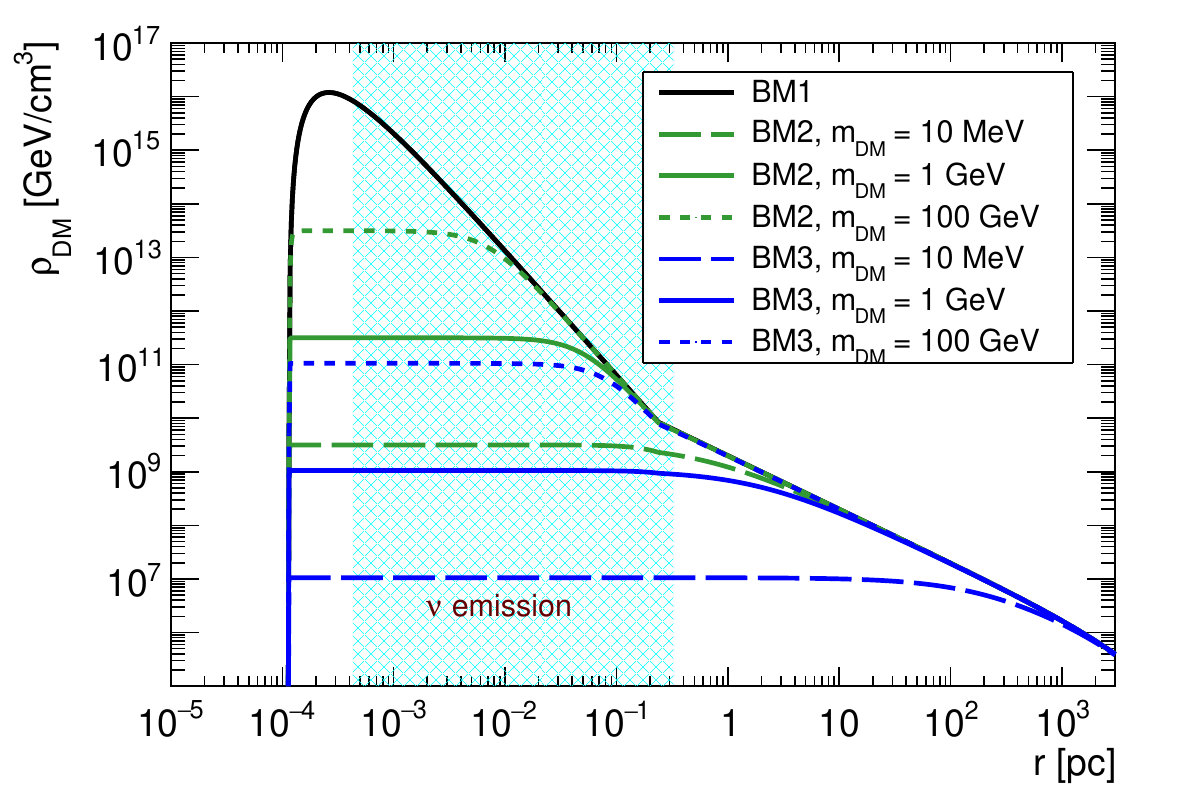}
\caption{\label{fig:spike_1} DM distribution around the black hole of TXS~0506+056, for the different considered benchmark models with $\rho_0 = 2 \times 10^5$ GeV/cm$^3$. In the left figure, the dark matter mass is assumed to be $m_{DM} = 1$~GeV. In light blue we show the region where neutrinos are likely emitted, see the text for details.}
\end{figure}
In Figure~\ref{fig:spike_1} we display the DM density profile of TXS~0506+056, for each of the benchmark models. On the left panel, we show $\gamma_{\textrm{sp}}=7/3$ ($3/2$) in black (red), with the previously mentioned values of $\langle\sigma v\rangle_{\textrm{ann}}$ in solid, dotted and dashed lines, respectively. It is clear that scenarios BM1 and BM1$'$, where $\langle\sigma v\rangle_{\textrm{ann}}$ vanishes, have a very cuspy profile, while scenarios BM3 and BM3$'$, where the annihilation is largest, show an inner core-like structure. On the right panel, we compare the profile of benchmarks BM1 - BM3 for different values of $m_{DM}$. For BM2 and BM3, where $\langle\sigma v\rangle_{\textrm{ann}}\neq0$, we find that the peak of the density profile is more suppressed when the DM mass is small.

The probability for neutrinos to scatter from DM in the spike of course depends on the amount of DM the neutrinos encounter on their path, which is encoded on the column density $\Sigma_{DM}$. To this end, we define the accumulated column density:
\begin{equation} \label{eq:column}
{\tilde\Sigma}_{DM}(r) = \int^{r}_{R_{\textrm{em}}} dr'\, \rho_{DM} (r')~,
\end{equation}
where $R_{\textrm{em}}$ is the location where the neutrinos are produced, measured from the center of the black hole. The column density for the DM density profile then follows $\Sigma_{DM}=\tilde\Sigma_{DM}(r\to\infty)$.
Contributions from the intergalactic DM and the Milky Way halo to $\Sigma_{DM}$ are about five orders of magnitude lower than the spike and TXS~halo contribution~\cite{Choi_2019}, so in the following are disregarded.

For clarity, in Table~\ref{tab:BM} we show the column density values obtained with the six different benchmark models, considering $m_{DM}=1$ GeV and $R_{\textrm{em}} =R_{BLR}=0.0227$ pc, the broad-line region (BLR), where the neutrinos are likely to be produced~\cite{Padovani_2019}. As expected, for a fixed $\gamma_{\textrm{sp}}$, the column density decreases with increasing $\langle\sigma v\rangle_{\textrm{ann}}$. Our results are consistent with Figure~\ref{fig:spike_1}, in the sense that larger density profiles correspond to larger values of $\Sigma_{DM}$. Notice that scenarios BM3 and BM3$'$ have practically identical profiles, regardless of the different $\gamma_{\textrm{sp}}$, leading to the same $\Sigma_{DM}$ in both cases.
\begin{table}
\centering
\begin{tabular}{c | c c c c c c}
\hline
Model & BM1 & BM2 & BM3 & BM1$'$ & BM2$'$ & BM3$'$ \\
\hline
$\gamma_{\textrm{sp}}$ & $7/3$ & $7/3$ & $7/3$ & $3/2$ & $3/2$ & $3/2$ \\
\hline
$\langle \sigma v \rangle_{\textrm{ann}}$ & $0$  & $0.01$  & $3$  & $0$  & $0.01$ & $3$ \\
\hline
$\Sigma_{DM}$ & $16.07$ & $10.14$ & $3.78$ & $8.70$ & $8.09$ & $3.78$ \\
\hline
\end{tabular}
\caption{\label{tab:BM} Benchmark models consider in this work, with $\langle \sigma v \rangle_{\textrm{ann}}$ in units of $10^{-26}$ cm$^3/$s. We also include the column density $\Sigma_{DM}$, in units of $10^{28}$ GeV/cm$^2$, calculated taking $R_{\textrm{em}} = R_{BLR}$ and $m_{DM}=1$ GeV.}
\end{table}

To finalize this Section, let us comment on the values of $R_{\textrm{em}}$ predicted by the models of Xue~\textit{et al.}~\cite{Xue_2021}, Wang~\textit{et al.}~\cite{Wang_2022} and Yang~\textit{et al.}~\cite{Yang:2024bsf}, described earlier. We follow the treatment made in~\cite{Cline_2022} and estimate the emission radius as $R_{\textrm{em}} = R'\, \delta$, where $R'$ is the comoving size of the emission region and $\delta$ is the Doppler factor.
Then, for the jet-cloud interaction model from Wang~\textit{et al.}~\cite{Wang_2022}, we obtain $R_{\textrm{em}} = 10^{18}\,{\textrm{cm}}\approx 0.3$~pc. For the two-zone model of Xue~\textit{et al.}~\cite{Xue_2021}, we find $R_{\textrm{em}} = 6 \times 10^{15}\,{\textrm{cm}}\approx 0.002$~pc.
For the accretion flow neutrinos model from Yang~\textit{et al.}~\cite{Yang:2024bsf}, we get $R_{\textrm{em}} \approx 1.3 \times 10^{15}\,{\textrm{cm}}\approx 4 \times 10^{-4}$~pc.
These values define the neutrino emission region, shown in light blue in Figure~\ref{fig:spike_1}. From this calculation, one finds that the flux from Wang~\textit{et al.}~\cite{Wang_2022} has $R_{\textrm{em}}\approx R_{\textrm{sp}}$, so their flux will not be subject to the spike. This does not mean that these neutrinos shall not be subject to interactions with DM, but that one should expect a relatively small $\Sigma_{DM}$. In contrast, the emission region for the fluxes by Xue~\textit{et al.}~\cite{Xue_2021} and  Yang~\textit{et al.}~\cite{Yang:2024bsf} lies closes to the peak of the spike, implying a large value of $\Sigma_{DM}$.

\section{Neutrino attenuation by DM} \label{sec:nudm}

Having defined the neutrino fluxes to be considered in our study, as well as the DM distribution around the SMBH, we now turn to describe the neutrino flux attenuation due to their scattering with DM along the journey to the detector. Assuming that neutrino - DM interactions are flavour-universal, this can be described by a form of the Boltzmann equation known as the cascade equation~\cite{Cline_2023, Cline_2022}:
\begin{equation} \label{eq:cascade}
\frac{d \Phi}{d \tau}(E_{\nu}, \tau) = - \sigma_{\nu DM}(E_{\nu}) \Phi(E_{\nu}, \tau) + \int_{E_{\nu}}^{\infty} d E'_{\nu} \frac{d \sigma_{\nu DM}}{d E_{\nu}} (E'_{\nu} \to E_{\nu}) \Phi(E'_{\nu}, \tau)~,
\end{equation}
where $\tau = \tilde\Sigma_{DM}(r) /m_{DM}$ is proportional to the accumulated column density from $R_{\textrm{em}}$ up to a point $r$. Here, $\Phi(E_{\nu}, \tau)$ is the neutrino flux after traversing a distance corresponding to a column density equal to $\tau\,m_{DM}$. On Eq.~\eqref{eq:cascade}, the first term on the right hand side describes the neutrino loss in the beam, depending directly on $\sigma_{\nu DM}$. Furthermore, the second term represents the redistribution of energy due to the neutrino - DM interaction, where an initial neutrino energy $E'_\nu$ is decreased to the observed energy $E_\nu$ by the $d\sigma_{\nu DM}/dE_\nu$ factor.

To solve this equation we must make an assumption about the energy dependence of both $\sigma_{\nu DM}$ and $d\sigma_{\nu DM}/dE_\nu$. In order to illustrate how the bound on the cross-section is placed, we first assume an energy independent cross-section, $\sigma_{\nu DM}=\sigma_0$, and neglect the energy redistribution term in Eq.~\eqref{eq:cascade}. This leads to a flux $\Phi_{\textrm{obs}}$ arriving to IceCube with an exponential attenuation:
\begin{equation}
\Phi_{\textrm{obs}}(E_\nu) = \Phi_{\textrm{em}}(E_\nu)\, e^{-\mu}
\end{equation}
where $\mu = \sigma_0 \Sigma_{DM} / m_{DM}$ and $\Phi_{\textrm{em}}(E_\nu)=\Phi(E_{\nu}, \tau=0)$ corresponds to the emitted neutrino flux. Then, using Eq.~\eqref{eq:nevents}, we can relate the parameter $\mu$ with the number of predicted and observed events:
\begin{equation}
\mu = \ln{\frac{N_{\textrm{pred}}}{N_{\textrm{obs}}}}
\end{equation}
where
\begin{equation}
N_{\textrm{pred}} = t_{\textrm{obs}} \int d E_{\nu}\, \Phi_{\textrm{em}} (E_{\nu})\, A_{\textrm{eff}} (E_{\nu}) \qquad
N_{\textrm{obs}} = t_{\textrm{obs}} \int d E_{\nu}\, \Phi_{\textrm{obs}} (E_{\nu})\, A_{\textrm{eff}} (E_{\nu})
\end{equation}
For this example, we take $\Phi_{\textrm{em}}$ to be the IceCube flux, as shown in Eq.~\eqref{eq:flux_ic1}, so $N_{\textrm{pred}} \approx 15$. Requiring that the observed events lie above the 90\% C.L. lower limit by IceCube ($N_{\textrm{obs}}\geq6.55$), we obtain $\mu\lesssim0.83$, so the energy independent cross section must satisfy:
\begin{equation}
\sigma_{0} \lesssim 0.83 \frac{m_{DM}}{\Sigma_{DM}}.
\end{equation}
In this case, the limit on the cross-section depends linearly on the ratio between DM mass and column density. From Table~\ref{tab:BM}, which assumes $m_{DM}=1$~GeV, we see that the limit ranges from $5.2$ to $22\times10^{-30}$~cm$^2$.

Let us now proceed with our full analysis. In the following, we take a cross section $\sigma_{\nu DM}$ with a power-law dependence with neutrino energy
\begin{equation} \label{eq:xsec}
\sigma_{\nu DM} (E_{\nu}) = \sigma_0 \left( \frac{E_{\nu}}{E_0} \right)^n, 
\end{equation}
with a reference energy $E_0 = 100$~TeV. For the scaling, we consider $n = 1, 0, -1, -2$, as motivated by simplified models such as those discussed in Appendix C of \mbox{\cite{Arg_elles_2017}}.\footnote{The scaling index $n$ is derived by taking the high- and low-energy limits of various cross sections. The cases $n=-1$ and $n=-2$ correspond, respectively, to models with fermionic and scalar dark matter, both with scalar mediators, in the limit of very high neutrino energy. A more detailed discussion on simplified models and their impact on neutrino flux attenuation will be presented in an upcoming paper.}
Moreover, for the differential cross section we consider a scattering isotropic in the center of mass frame, and approximate:
\begin{equation} \label{eq:dxsec}
\frac{d \sigma_{\nu DM}}{d E_{\nu}} (E'_{\nu} \to E_{\nu}) \approx \frac{\sigma_{\nu DM}(E'_{\nu})}{E'_{\nu}} = \frac{ \sigma_0 }{E_0} \left( \frac{E'_{\nu}}{E_0} \right)^{n-1}.
\end{equation}

This assumption for the neutrino - DM cross-section allows us to solve the cascade equation, following the algorithm presented in~\cite{Vincent_2017}. For this, we evaluate the flux at specific values of energy $E_i$, such that $\Phi_i(\tau)\equiv\Phi(E_i,\,\tau)$. The chosen $E_i$ are logarithmically spaced, that is, $E_i=10^{x_i}$, with constant $\Delta x$. This allows us to discretize Eq.~\eqref{eq:cascade}, taking the form:
\begin{equation} \label{eq:casc_dis_pro}
\frac{d \Phi_i}{dy} = \mu \sum_{j=i}^{N}\left( -\left( \frac{E_i}{E_0} \right)^n \delta_{ij} + \Delta x \ln 10 \left( \frac{E_j}{E_0} \right)^n  \right)\Phi_j
\end{equation}
where we have defined a dimensionless evolution parameter $y = (m_{DM} / \Sigma _{DM}) \tau \in [0,1]$. With this, Eq.~\eqref{eq:casc_dis_pro} can be written in terms of a matrix $M$, such that:
\begin{equation}
\frac{d \vec{\Phi}}{dy} = \mu M \vec{\Phi}~.
\end{equation} 
Since the equation is linear, the eigenvectors $\hat{\phi}_i$ of $M$ satisfy the differential equation $\hat{\phi}'_i= \mu\, \lambda_i\, \hat{\phi}_i$, where $\lambda_i$ are the corresponding eigenvalues.
With the $\hat{\phi}_i$ forming a complete basis, the solution of Eq.~\eqref{eq:casc_dis_pro} is
\begin{equation}
\vec{\Phi}(y) = \sum  c_i\, \hat{\phi}_i\, e^{\mu\, \lambda_i\, y}~,
\end{equation}
where the coefficients $c_i$ are determined by the initial neutrino flux, at $y=0$. The observed flux $\Phi_{\textrm{obs}}(E_\nu)$ is an interpolation of the solved $\Phi_i(y=1)$, which can be used to calculate $N_{\textrm{obs}}$ in Eq.~(\ref{eq:nevents}). We also checked we got the same results when solving Eq.~\eqref{eq:casc_dis_pro}, by evolving the initial flux with small increments in $y$, from $y=0$ to $y=1$.

\begin{figure}[tb]
\centering
\begin{subfigure}[b]{0.49\linewidth}
	\includegraphics[width=1\textwidth]{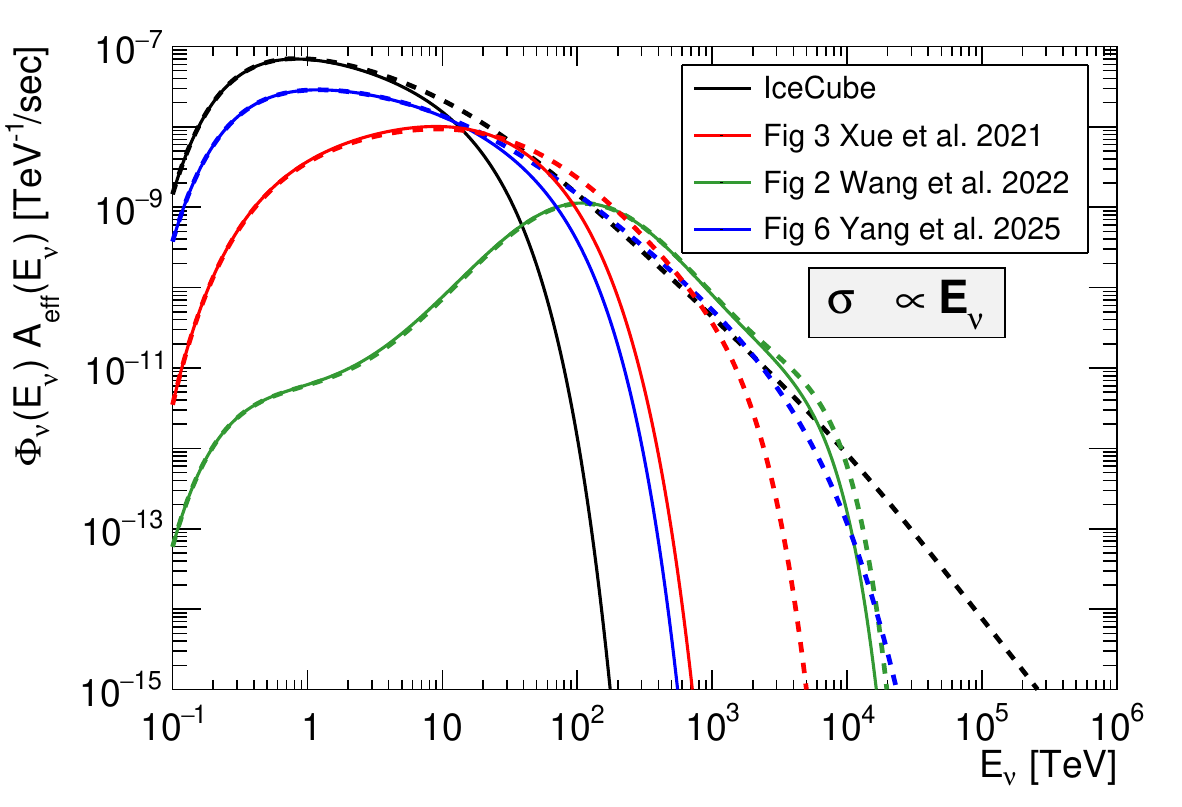}
	\caption{}
	\label{fig:flux_out_lin}
\end{subfigure}
\begin{subfigure}[b]{0.49\linewidth}
	\includegraphics[width=1\textwidth]{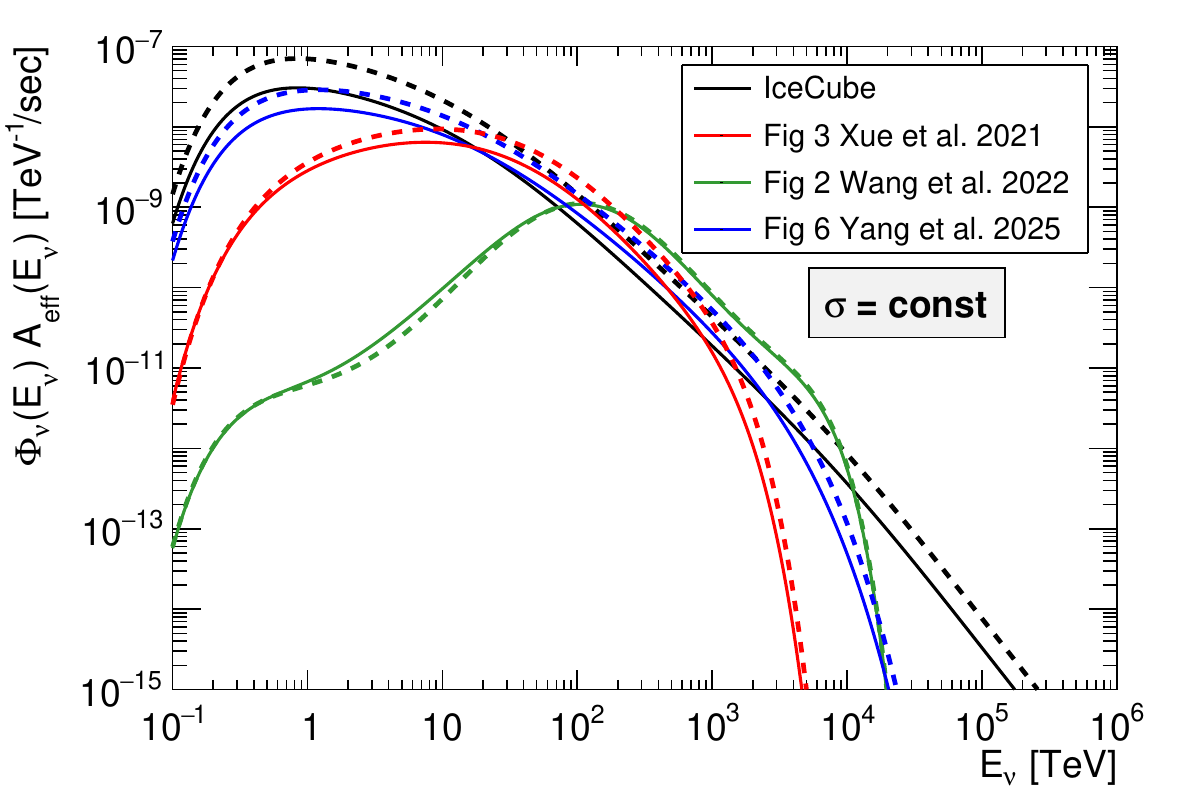}
	\caption{}
	\label{fig:flux_out0}
\end{subfigure}
\begin{subfigure}[b]{0.49\linewidth}
	\includegraphics[width=1\textwidth]{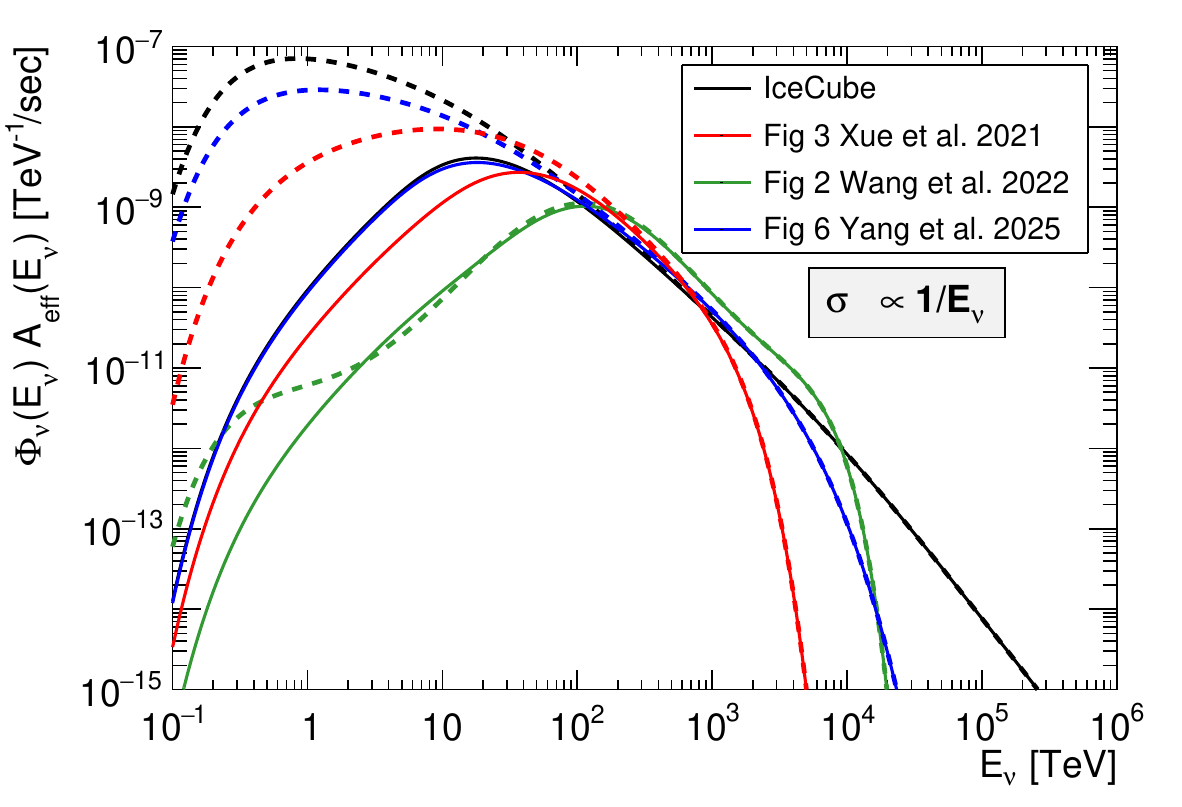}
	\caption{}
	\label{fig:flux_out_inv}
\end{subfigure}
\begin{subfigure}[b]{0.49\linewidth}
	\includegraphics[width=1\textwidth]{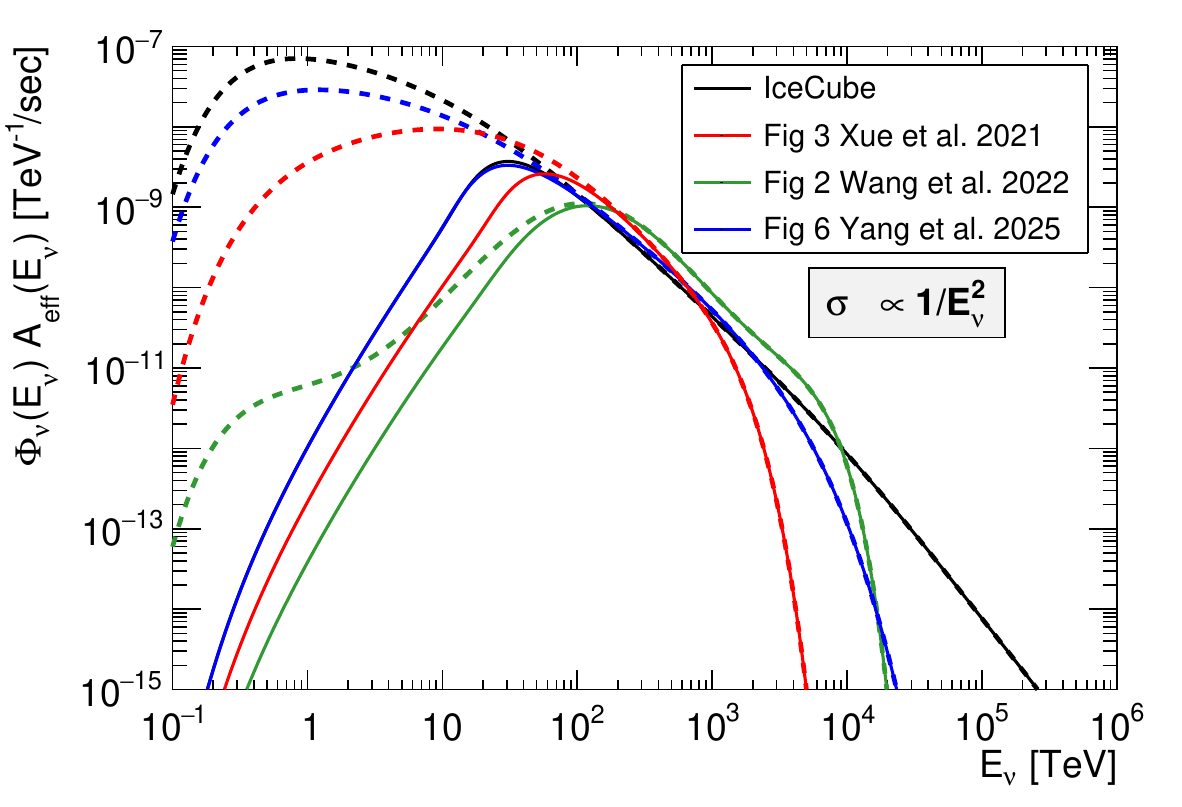}
	\caption{}
	\label{fig:flux_out_inv2}
\end{subfigure}
\caption{Neutrino spectrum at IceCube, allowing interactions with DM. The solid lines show the spectra for all the considered fluxes with a cross section $\sigma_{\nu DM} \propto E_{\nu}^n$ for a) $n=1$, b) $n=0$, c) $n=-1$ and d) $n=-2$. All curves are consistent with $N_{\textrm{obs}} = 6.55$. The spectrum representing the case where $\sigma_{\nu DM}=0$, also presented on the right panel of Figure~\ref{fig:aeff}, is shown in dashed lines.}
\label{fig:flux_out}
\end{figure}

\begin{table}
\centering
\begin{tabular}{l | c c c c}
\hline 
\multirow{2}{10em}{Reference flux $\Phi_{\nu}$} & \multicolumn{4}{c}{$\mu_{\textrm{max}}$} \\
& $n=1$ & $n=0$ & $n=-1$ & $n=-2$ \\
\hline 
IceCube~\cite{Aartsen_2018b}  & $9.32$ & $1.54$ & $0.318$ & $0.0801$\\
\hline 
Xue \textit{et al.}~\cite{Xue_2021} & $1.89$ & $1.13$ & $0.505$ & $0.228$ \\ 
\hline 
Wang \textit{et al.}~\cite{Wang_2022} & $0.0157$ & $0.0977$ & $0.173$ & $0.134$ \\ 
\hline 
Yang \textit{et al.}~\cite{Yang:2024bsf} & $2.62$ & $1.08$ & $0.267$ & $0.0664$ \\ 
\hline 
\end{tabular}
\caption{\label{tab:mu_IC} Maximum allowed value for the attenuation parameter $\mu = \sigma_0 \Sigma_{DM}/m_{DM}$ for the different neutrino flux models, assuming $\sigma_{\nu DM} = \sigma_0 (E_\nu/E_0)^n$.}
\end{table}
Constraints on $\mu$ are then placed by requiring $N_{\textrm{obs}}\geq6.55$, as done earlier. The largest allowed values, $\mu_{\textrm{max}}$, are reported in Table~\ref{tab:mu_IC}, for $n=1,0,-1,-2$ and the four assumed fluxes. For the IceCube flux in Eq.~\eqref{eq:flux_ic1}, constraints are weakest for $\sigma_{\nu DM}\propto (E_\nu/E_0)$, ($n=1$), and strongest for $\sigma_{\nu DM}\propto (E_\nu/E_0)^{-2}$, ($n=-2$). This can be understood by analyzing how each cross-section affects the low-energy and high-energy parts of the neutrino spectrum at IceCube, illustrated in black lines on Figure~\ref{fig:flux_out}. For $n=1$, shown in Figure~\ref{fig:flux_out_lin}, the high-energy tail of the flux is strongly suppressed, particularly for energies above 10~TeV. However, since the peak of the spectrum lies at $E_\nu\approx 1$~TeV, losing the high-energy tail affects $N_{\textrm{obs}}$ only marginally, allowing then large values for $\mu_{\textrm{max}}$. In contrast, as seen on Figure~\ref{fig:flux_out_inv2}, setting $n=-2$ affects heavily the low-energy part of the spectrum. The small value of $\mu_{\textrm{max}}$ in this case already decreases the peak by more than one order of magnitude, shifting it to $E_\nu\approx30$~TeV, thus suppressing $N_{\textrm{obs}}$ down to $6.55$ events.

Still concentrating on the IceCube flux, the cases with $n=0$ and $n=-1$, on black lines in Figs.~\ref{fig:flux_out0} and~\ref{fig:flux_out_inv} respectively, show in-between scenarios. The energy-independent cross-section affects the whole spectrum without shifting the position of the peak, with the maximum value diminished by a factor $\sim2$. Since this time the low-energy part of the spectrum is affected, $\mu_{\textrm{max}}$ cannot be as large as the one for $n=1$. In contrast, when the cross-section is proportional to the inverse of $E_\nu$, the peak is suppressed and shifted, similar to the $n=-2$ scenario, implying a smaller $\mu_{\textrm{max}}$.

Table~\ref{tab:mu_IC} and Figure~\ref{fig:flux_out} also show results for the three blazar models~\cite{Xue_2021,Wang_2022,Yang:2024bsf}. Before giving details, it is worth noting that for $\sigma_{\nu DM}=0$ the models by Xue~\textit{et al.}~\cite{Xue_2021} and Yang~\textit{et al.}~\cite{Yang:2024bsf} predict over $50\%$ more events than the one from Wang~\textit{et al.}~\cite{Wang_2022}, so it is natural to expect the former two to be less constraining than the latter. Furthermore, the prediction from Wang~\textit{et al.}~\cite{Wang_2022} in this case is very close to our threshold, $N_{\textrm{obs}}\geq6.55$, so for this model we should expect small modifications to the spectrum shown on the right panel of Figure~\ref{fig:aeff}.

As can be seen in Table~\ref{tab:mu_IC}, for the fluxes by Yang~\textit{et al.}~\cite{Yang:2024bsf} and Xue~\textit{et al.}~\cite{Xue_2021} the bounds on $\mu$ becomes more constraining as one moves from $n=1$ to $n=-2$. The underlying reason for this is similar to that for the IceCube flux, to pass the $N_{\textrm{obs}}$ threshold these models rely mostly on low- and medium-energy events (around 1 and 10~TeV, respectively), as can be corroborated in Figure~\ref{fig:flux_out}. The $n=1$ interactions affect mainly the tail of the spectrum, so these models can afford a larger $\mu$, while the $n=-2$ interactions suppress the low and medium energies, forcing $\mu$ to lower values. Notice that, except for $n=1$, the flux by Yang~\textit{et al.}~\cite{Yang:2024bsf} puts stronger bounds on $\mu$ than the one by Xue~\textit{et al.}~\cite{Xue_2021}, which is indicative of the higher importance of the lowest energy part of the spectrum for this model.

Let us now turn to the model by Wang~\textit{et al.}~\cite{Wang_2022} which, as mentioned previously, already has a small number of events in absence of interactions with DM, leading to $\mu_{\textrm{max}}$ values that are usually small. However, contrary to the previous cases, it is less constraining as one moves from $n=1$ to $n=-1$. As one can see in Figure~\ref{fig:flux_out}, this happens because the flux presents a hard spectra, peaking at $\sim100$~TeV, with a small contribution to $N_{\textrm{obs}}$ coming from low-energy events. Thus, $n=-1$ tends to affect a sector of the spectrum that does not contribute much to the final number of events. Interestingly, $n=-2$ is less constrained than $n=-1$, here the suppression of low-energy events is so strong that it starts affecting the 100~TeV peak.

We can also see in Figure~\ref{fig:flux_out}, for the cases $n=0$ and $n=-1$, that the observed flux becomes slightly larger than the initial flux for $E_{\nu} \sim 10$ TeV. This is due to the effect of the second term in the cascade equation \eqref{eq:casc_dis_pro}, which enhances the low-energy neutrino flux at the expense of the high-energy flux. Given the hard spectrum of the Wang flux, this effect becomes more pronounced.

\begin{figure}[tb]
\centering
\begin{subfigure}[b]{0.49\linewidth}
	\includegraphics[width=1\textwidth]{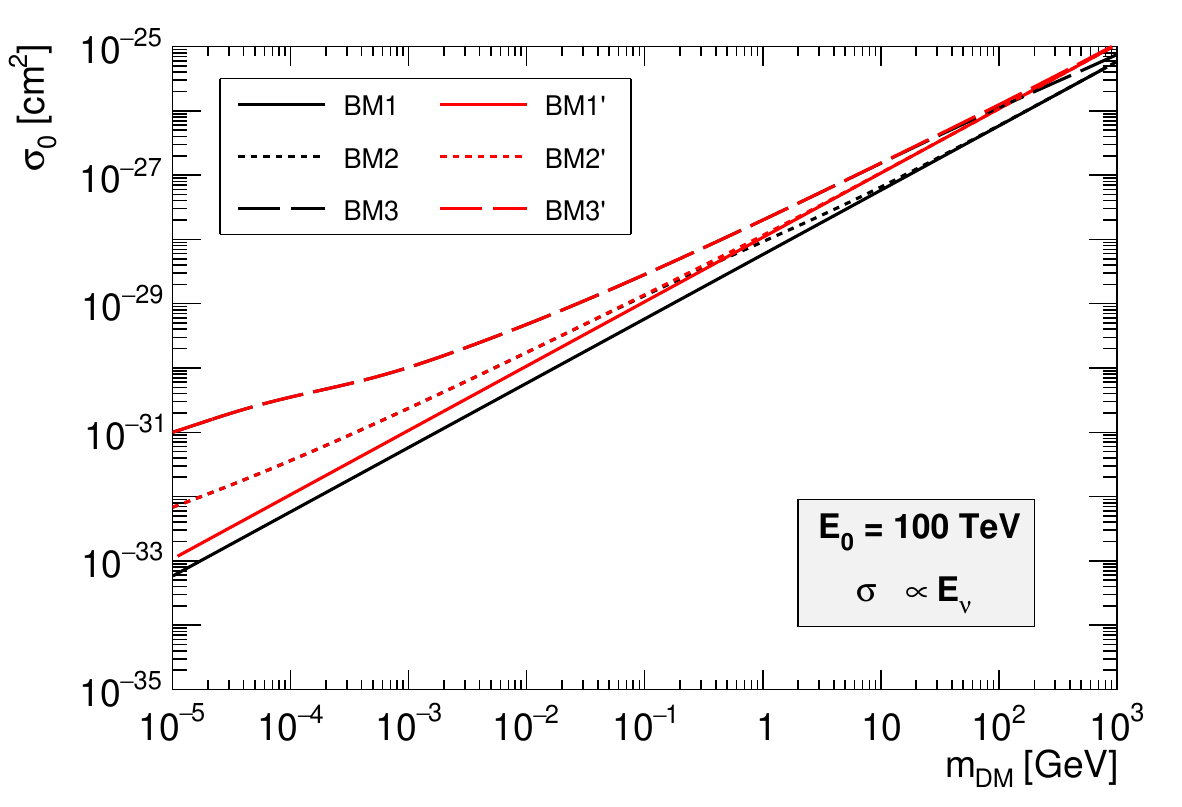}
	\caption{}
	\label{fig:xsec_elin}
\end{subfigure}
\begin{subfigure}[b]{0.49\linewidth}
	\includegraphics[width=1\textwidth]{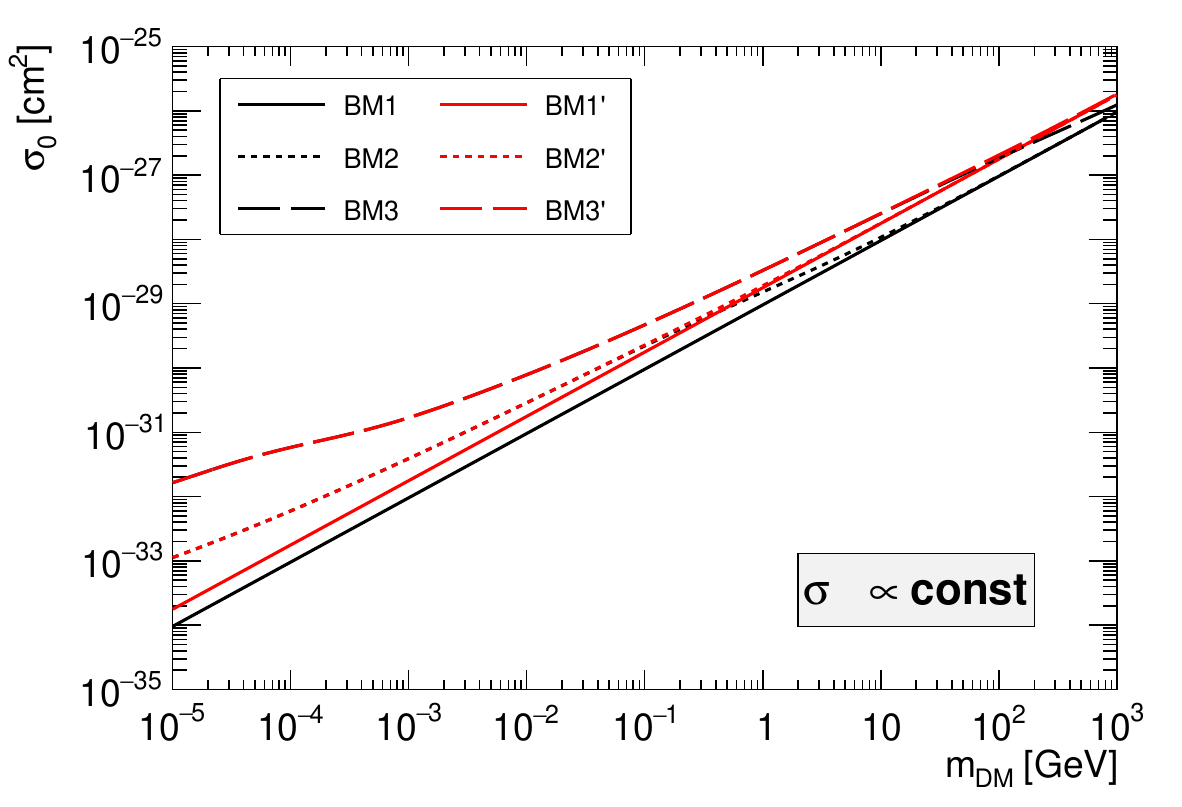}
	\caption{}
	\label{fig:xsec_ein}
\end{subfigure}
\begin{subfigure}[b]{0.49\linewidth}
	\includegraphics[width=1\textwidth]{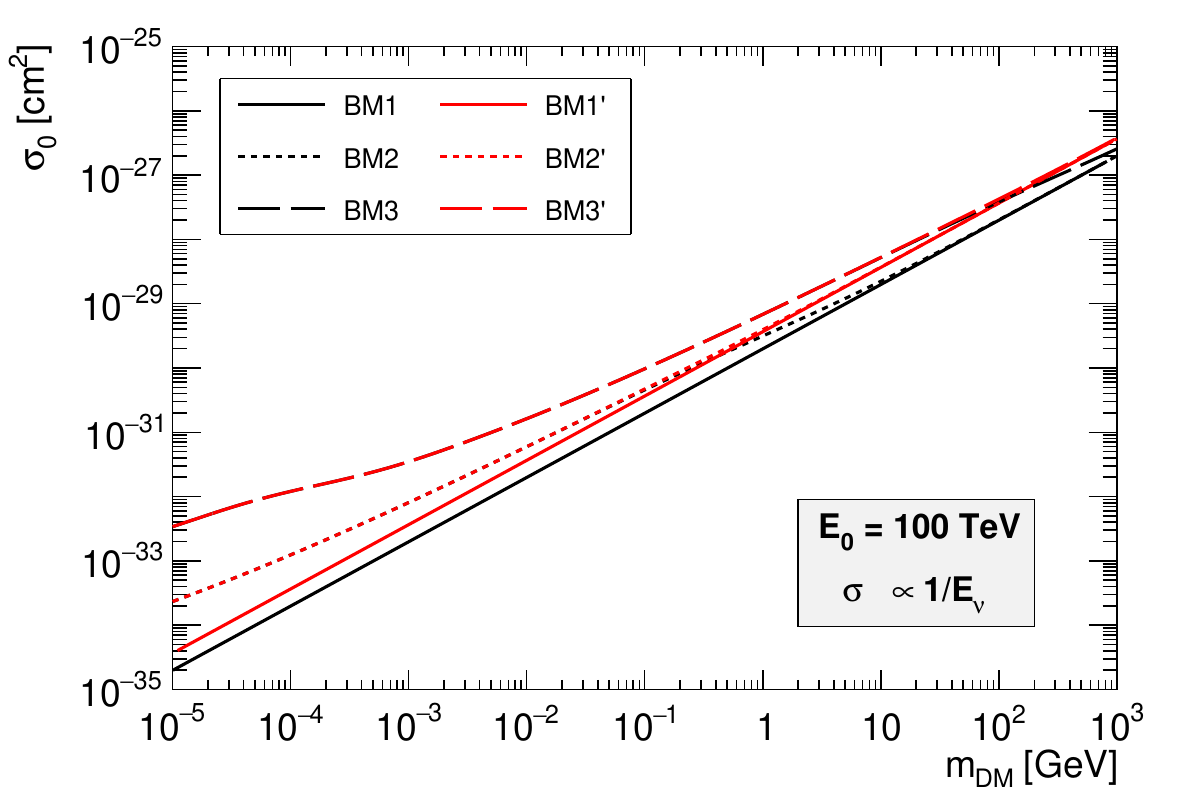}
	\caption{}
	\label{fig:xsec_einv}
\end{subfigure}
\begin{subfigure}[b]{0.49\linewidth}
	\includegraphics[width=1\textwidth]{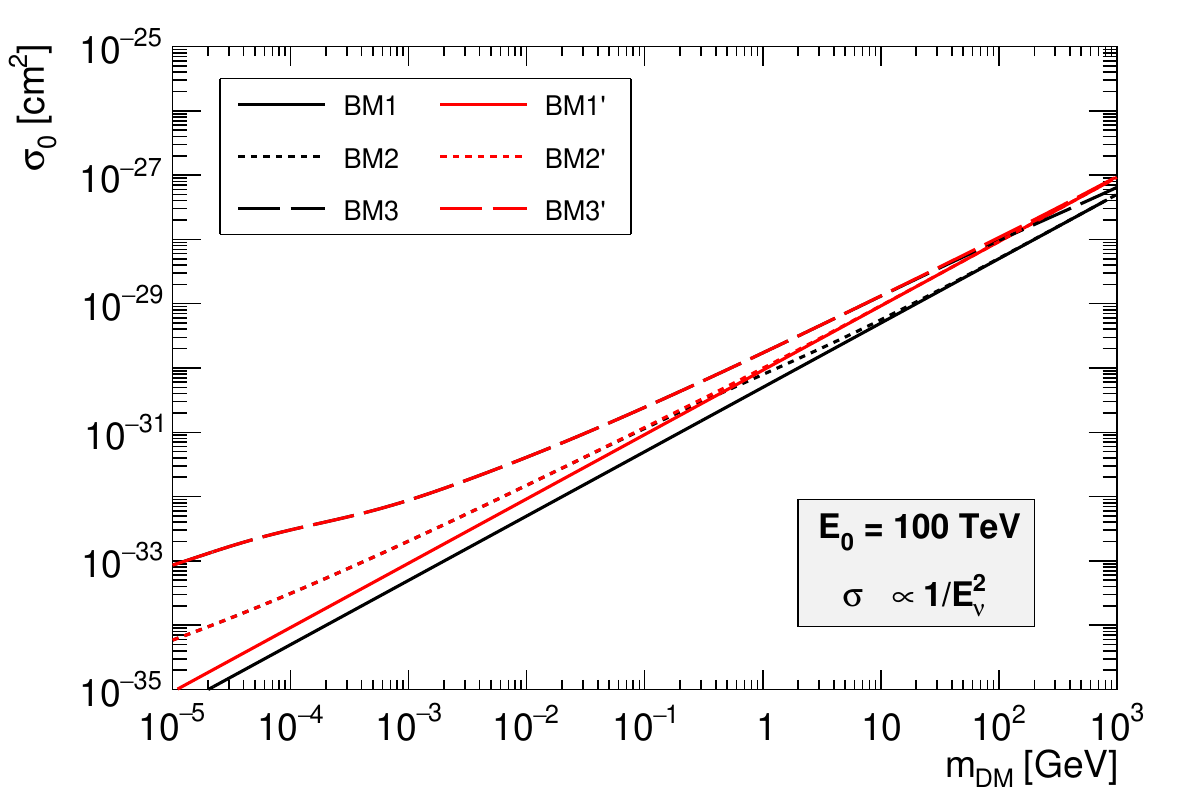}
	\caption{}
	\label{fig:xsec_einv2}
\end{subfigure}
\caption{90\% C.L. bounds on DM-neutrino cross section obtained with $R_{\textrm{em}} = R_{BLR} $ and flux given by Eq.~\eqref{eq:flux_ic1}, considering (a) $\sigma \propto E_{\nu}$ ($n=1$), (b) $\sigma = \sigma_0$ ($n=0$), (c) $\sigma \propto 1/E_{\nu}$ ($n=-1$) and (d) $\sigma \propto 1/E^2_{\nu}$ ($n=-2$).}
\label{fig:xsec_IC}
\end{figure}
It is important to notice that these results are independent of any assumption taken regarding the DM mass and spike profile, as all this information is contained within $\mu$.\footnote{Note that the bound on $\mu$ depends on $E_0$, the chosen reference energy for $\sigma_0$. In spite of this, all results shown in Figs.~\ref{fig:flux_out} do not depend on this choice.} Nevertheless, the latter become relevant when translating the limit on $\mu$ to a limit on $\sigma_0$, the reference cross-section at $E_\nu=E_0=100$~TeV:
\begin{equation}
\label{eq:sigma0}
\sigma_{0} \leq \mu_{\textrm{max}}\, \frac{m_{DM}}{\Sigma_{DM}}~.
\end{equation}
Such a bound is shown in Figure~\ref{fig:xsec_IC} as a function of $m_{DM}$, assuming the IceCube flux and setting the neutrino emission region $R_{\textrm{em}}=R_{BLR}$. Each panel shows one of the four assumed behaviors for the cross-section, and places constraints based on the benchmark models for the DM spike profile, as detailed in Table~\ref{tab:BM}.

Let us first focus on benchmarks BM1 and BM1$'$, shown on all panels in solid black and red lines, respectively. These two benchmarks assume no DM self-annihilation, meaning that they have the largest column densities, and thus place the strongest limits on $\sigma_0$. For $m_{DM}=10$~keV, the limits range from $\sim\mathcal O(10^{-33})$~cm$^2$ ($n=1$) to $\sim\mathcal O(10^{-36})$~cm$^2$ ($n=-2$), while for $m_{DM}=1$~TeV, the limits are weakened to $\sim\mathcal O(10^{-25})$~cm$^2$ ($n=1$) to $\sim\mathcal O(10^{-28})$~cm$^2$ ($n=-2$).

For benchmarks BM1 and BM1$'$  described above, the limits depend linearly on $m_{DM}$. However, this is no longer the case when allowing for DM self-annihilation, as the saturation density depends on the DM mass, see the right panel of Figure~\ref{fig:spike_1} and the discussion below Eq.~(\ref{eq:rho}). This means that $\Sigma_{DM}$ becomes a function not only of $\rho_{DM}(r)$ and $R_{\textrm{em}}$, but also of $m_{DM}$, with lower values of DM mass associated to a larger suppression of the column density. Such behavior is evident in Figure~\ref{fig:xsec_IC} where, for large DM mass, the bounds in benchmarks BM2$^{(\prime)}$ and BM3$^{(\prime)}$, shown in dotted and dashed lines, tend to coincide with those for benchmarks BM1$^{(\prime)}$, while for small DM mass they are much weaker. In particular, for $m_{DM}=1$~keV, the constraints on $\sigma_0$ in benchmarks BM3$^{(\prime)}$ are roughly two orders of magnitude weaker than those for BM1$^{(\prime)}$.

Finally, we address the bounds on $\sigma_0$ for the models in consideration~\cite{Xue_2021,Wang_2022, Yang:2024bsf}. As commented in Section~\ref{sec:dm_spike}, one can estimate $R_{\textrm{em}}$ in each case. The model by Wang~\textit{et al.}~\cite{Wang_2022} have the largest $R_{\textrm{em}}$, such that their emission region is practically outside the DM spike. In contrast, the flux by Yang~\textit{et al.}~\cite{Yang:2024bsf} has the smallest $R_{\textrm{em}}$, that is, it is generated deep within the spike. It is then to be expected that the column density of the former will be the smallest, and the latter will be largest. As an example, for BM1 we have $\Sigma_{DM}=(16.07,\,261,\,5.80,\,1318.7)\times10^{28}$~GeV/cm$^2$ for IceCube, Xue~\textit{et al.}~\cite{Xue_2021}, Wang~\textit{et al.}~\cite{Wang_2022} and Yang~\textit{et al.}~\cite{Yang:2024bsf} fluxes, accordingly. 

The value of $R_{\textrm{em}}$ is also connected to the sensitivity to changes in the DM density profile. As shown in Figure~\ref{fig:spike_1}, the density profile can change by the modification of $\gamma_{\textrm{sp}}$, or by the presence of DM self-annihilation, with this last feature depending on $m_{DM}$ and $\langle\sigma v\rangle_{\textrm{ann}}$. Furthermore, alterations due to self-annihilation are stronger for $r<R_{\textrm{sp}}$, that is, the spike is more sensitive to them than the NFW profile. Thus, the fluxes from Xue~\textit{et al.}~\cite{Xue_2021} and Yang~\textit{et al.}~\cite{Yang:2024bsf} should be most affected by changes in $\gamma_{\textrm{sp}}$, $\langle\sigma v\rangle_{\textrm{ann}}$ and $m_{DM}$, while those by Wang~\textit{et al.}~\cite{Wang_2022} should not be affected by $\gamma_{\textrm{sp}}$, and would be marginally sensitive to the other two parameters (apart from the explicit $m_{DM}$ dependence expected from Eq.~(\ref{eq:sigma0})).

\begin{figure}[tb]
\centering
\begin{subfigure}[b]{0.49\linewidth}
	\includegraphics[width=1\textwidth]{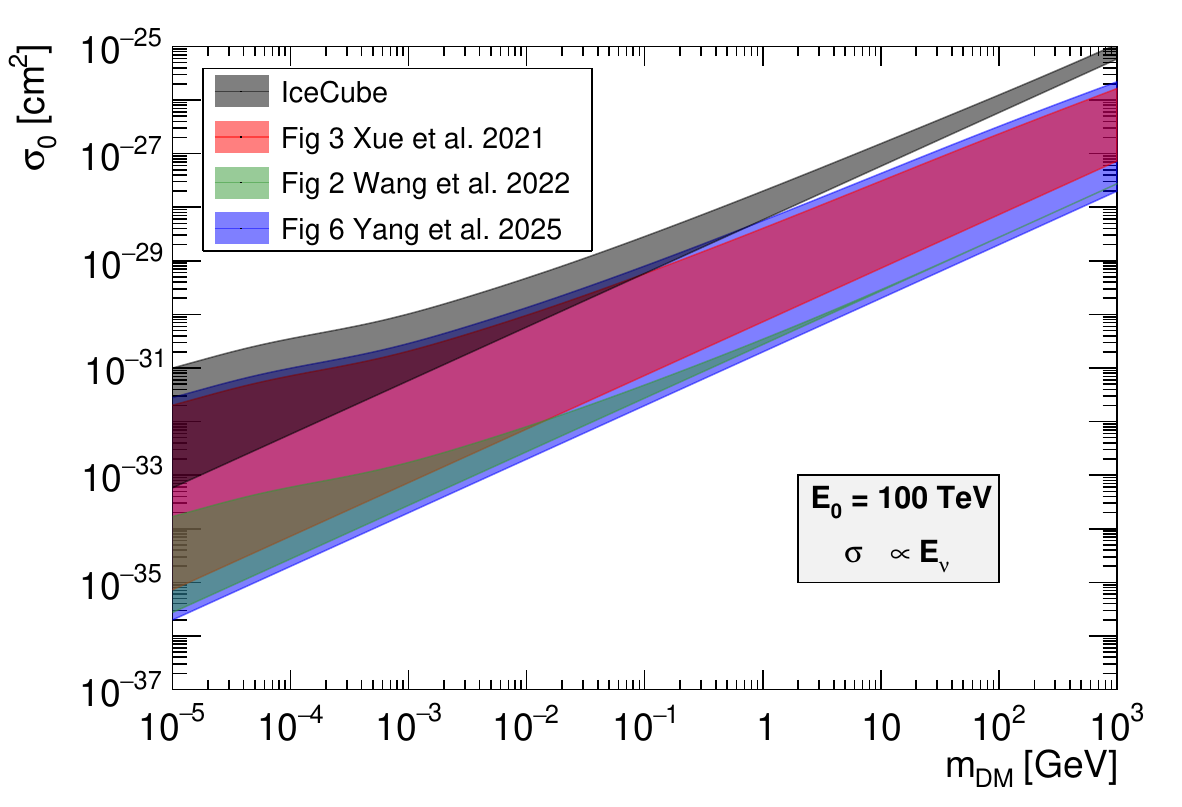}
	\caption{}
	\label{fig:reg_n1}
\end{subfigure}
\begin{subfigure}[b]{0.49\linewidth}
	\includegraphics[width=1\textwidth]{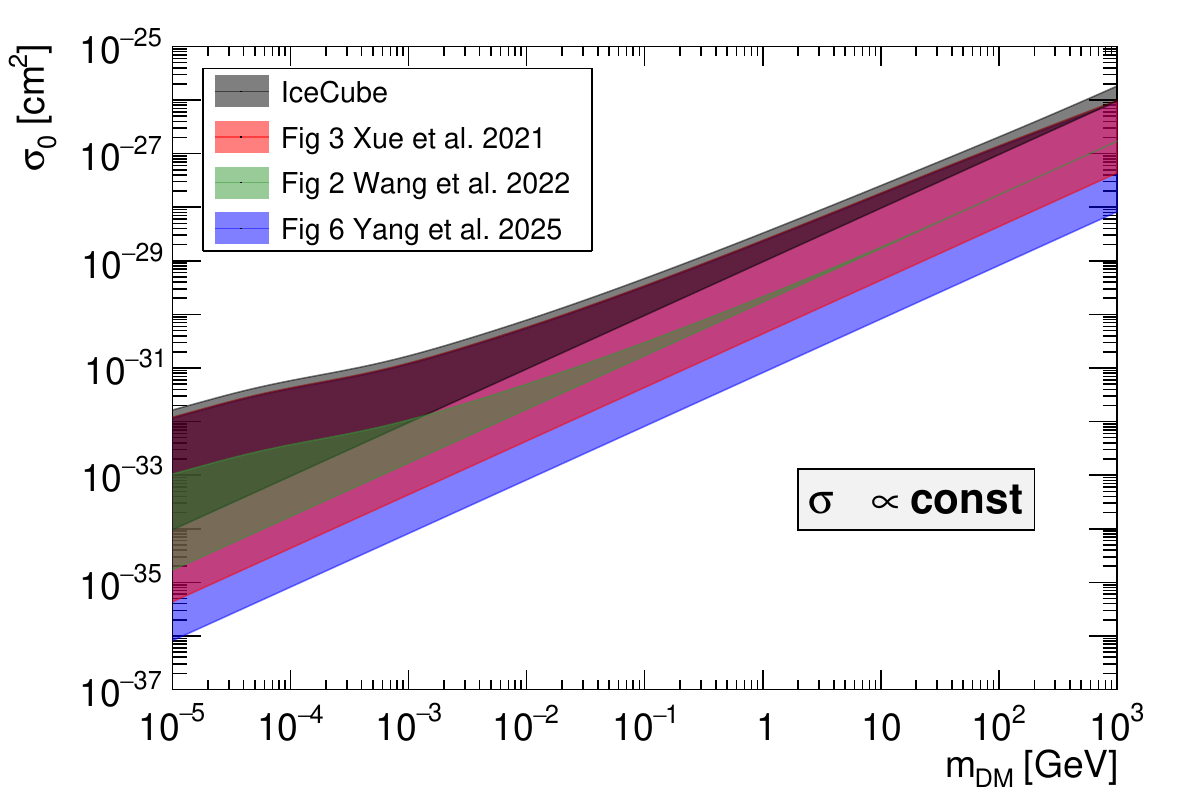}
	\caption{}
	\label{fig:reg_n0}
\end{subfigure}
\begin{subfigure}[b]{0.49\linewidth}
	\includegraphics[width=1\textwidth]{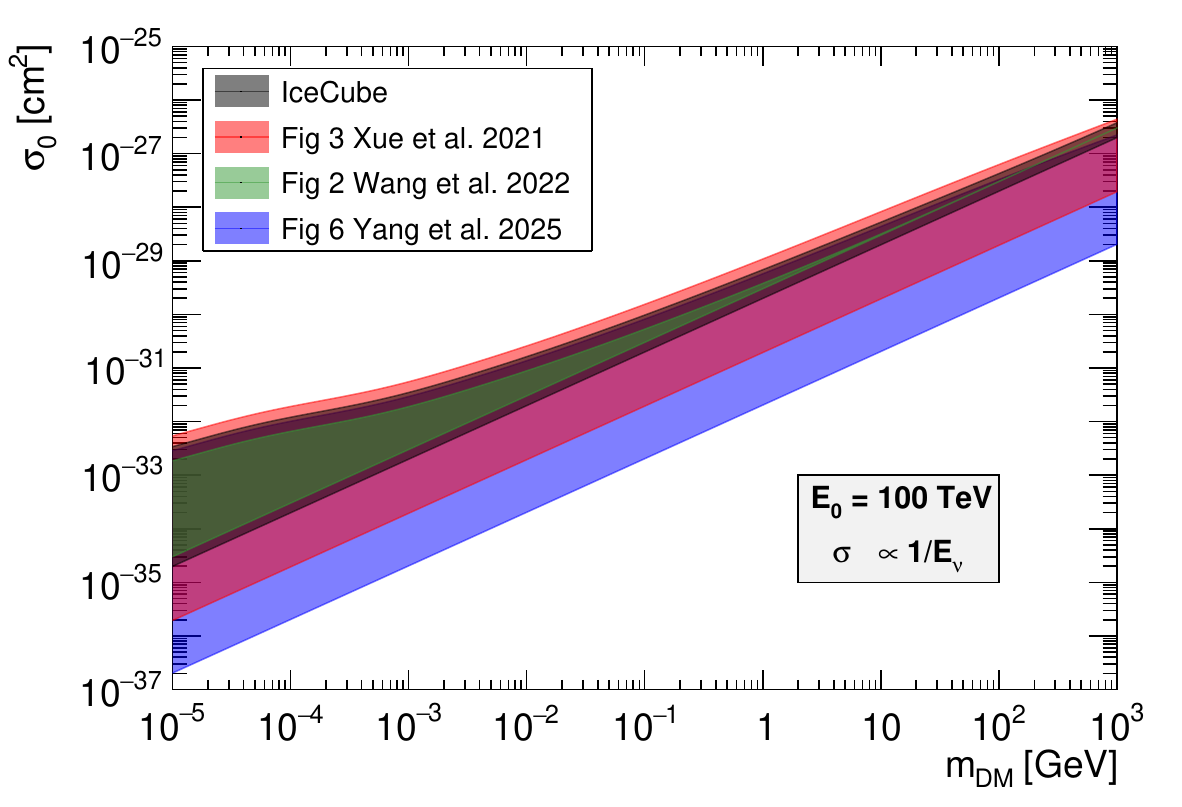}
	\caption{}
	\label{fig:reg_n-1}
\end{subfigure}
\begin{subfigure}[b]{0.49\linewidth}
	\includegraphics[width=1\textwidth]{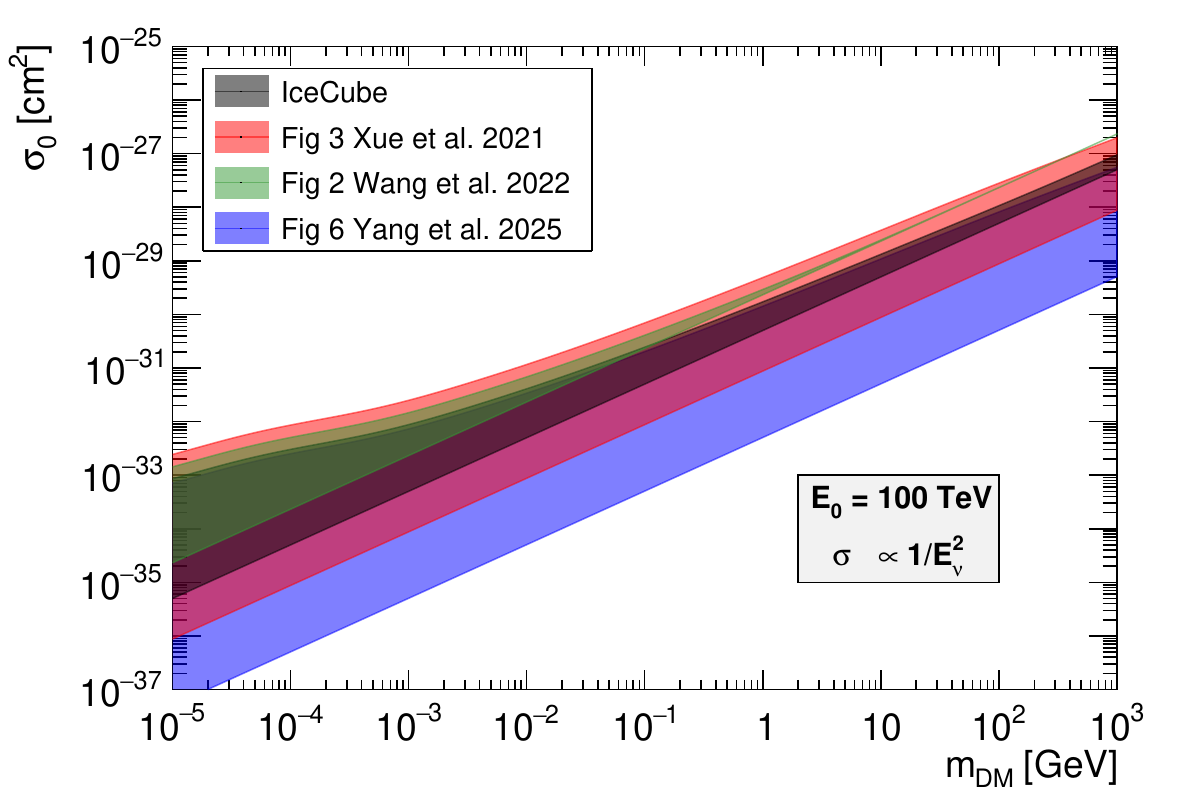}
	\caption{}
	\label{fig:reg_n-2}
\end{subfigure}
\caption{Range of bounds between BM1 and BM3' for each flux mode with a cross section $\sigma_{\nu DM} \propto E_{\nu}^n$ for a) $n=1$, b) $n=0$, c) $n=-1$ and d) $n=-2$.}
\label{fig:regions}
\end{figure}
These considerations imply that even though the attenuation parameter $\mu$ could be less constrained in a specific model compared to another one, a large column density could turn this into a stronger limit in terms of $\sigma_0$. This is shown in Figure~\ref{fig:regions}, where we plot bounds on $\sigma_0$ as a function of the DM mass. As before, we show results for a cross-section proportional to $(E_\nu/E_0)^n$, for $n=1,\,0,\,-1,\,-2$. The Figure shows a shaded region for each flux, indicating a range of bounds, obtained by comparing the tightest and loosest bounds, from BM1 and BM3$'$, respectively.

Let us focus on results for $m_{DM}=1$~TeV (the right hand side of all panels). Here, the DM density profile is least affected by self-annihilation, so the uncertainty in the bound for each model is the lowest. Consistent with our expectations, for this DM mass, the bound from Wang~\textit{et al.}~\cite{Wang_2022} have a very small uncertainty, followed by that from IceCube, and finally the ones from Xue~\textit{et al.}~\cite{Xue_2021} and Yang~\textit{et al.}~\cite{Yang:2024bsf}, where the uncertainty is relatively large, mainly due the variation in $\gamma_{sp}$. In contrast, for $m_{DM}=10$~keV (the left hand side of all panels) we have that even the NFW profile at $r>R_{\textrm{sp}}$ is affected by the variation in $\langle\sigma v\rangle_{\textrm{ann}}$, leading to a large uncertainty for the bounds from all models. Still, the limits from Yang~\textit{et al.}~\cite{Yang:2024bsf} always have the largest uncertainty over the whole DM mass range we have considered.

For $n=1$, Figure~\ref{fig:reg_n1}, the constraints placed assuming the flux from Yang~\textit{et al.}~\cite{Yang:2024bsf} are potentially the strongest, for all values of $m_{DM}$.
The weakest bounds, in turn, always come from the IceCube flux, for $m_{DM}\gtrsim100$~MeV. For lower masses, the maximum uncertainty in this case reaches somewhat over two orders of magnitude, also overlapping the bounds from Xue~\textit{et al.}~\cite{Xue_2021} and Yang~\textit{et al.}~\cite{Yang:2024bsf}. Furthermore, we find in this case that the range of bounds from Xue~\textit{et al.}~\cite{Xue_2021} and Wang~\textit{et al.}~\cite{Wang_2022} are always fully contained within the range from Yang~\textit{et al.}~\cite{Yang:2024bsf}. Overall, the constraints placed for $m_{DM}=10$~keV range from $\mathcal O(10^{-36})$ to $\mathcal O(10^{-31})$~cm$^2$, while those for $m_{DM}=1$~TeV go from $\mathcal O(10^{-28})$ to about $\mathcal O(10^{-25})$~cm$^{2}$.

For $n=0$, Figure~\ref{fig:reg_n0}, the strongest constraint is potentially placed by Yang~\textit{et al.}~\cite{Yang:2024bsf} over the whole mass range. However, the uncertainty is so large, that it encompasses the range of bounds placed by Wang~\textit{et al.}~\cite{Wang_2022}, the range by Xue~\textit{et al.}~\cite{Xue_2021}, and most of that by the IceCube flux. The latter also always potentially places the weakest bound. Overall, for $m_{DM}=10$~keV, the bounds span from $\mathcal O(10^{-36})$ to $\mathcal O(10^{-32})$~cm$^2$. For $m_{DM}=1$~TeV, the limits reach as low as $\mathcal O(10^{-28})$ and as high as $\mathcal O(10^{-26})$~cm$^2$.

For $n=-1$, Figure~\ref{fig:reg_n-1}, and $n=-2$, Figure~\ref{fig:reg_n-1}, the model by Yang~\textit{et al.}~\cite{Yang:2024bsf} always potentially provides the most stringent constraints. The second strongest constraints are given by the model of Xue~\textit{et al.}~\cite{Xue_2021}, though its large uncertainty also leads it to potentially giving the weakest constraints of all the models. The range from the latter model fully encompasses those from IceCube and Wang~\textit{et al.}~\cite{Wang_2022}, except for a small, narrow region above $\sim300$~GeV for $n=-2$ where Wang~\textit{et al.}~\cite{Wang_2022} gives much weaker bounds. It is also interesting to note that for $n=-1$ the range of constraints from Wang~\textit{et al.}~\cite{Wang_2022} is also fully contained within the range from IceCube. 

Overall, for $n=-1$ and $m_{DM}=1$~keV, the constraints go from $\mathcal O(10^{-37})$ to $\mathcal O(10^{-32})$~cm$^2$. For the same $n$ but $m_{DM}=1$~TeV, the bounds span $\mathcal O(10^{-29})$ to $\mathcal O(10^{-27})$~cm$^2$. For $n=-2$, we find that for $m_{DM}=10$~keV we have limits going from below $\mathcal O(10^{-37})$ up to $\mathcal O(10^{-33})$~cm$^2$. Furthermore, for $m_{DM}=1$~TeV, the bounds range from $\mathcal O(10^{-29})$ to $\mathcal O(10^{-27})$~cm$^2$.

To summarize, we find that the bounds on $\sigma_0$ depends very strongly on the neutrino emission model, the DM mass $m_{DM}$, the DM self-annihilation cross section $\langle\sigma v\rangle_{\textrm{ann}}$, and the energy scaling $n$ of the neutrino - DM cross-section. Without any further information on any of these inputs, the bound can vary from $\mathcal O(10^{-37})$ to $\mathcal O(10^{-25})$~cm$^2$. However, this situation will greatly improve once the DM mass is measured, reducing the overall uncertainty to within $2-5$ orders of magnitude.

As commented earlier, it is important to take into account that the flux from Wang~\textit{et al.}~\cite{Wang_2022} already have a low number of events when $\sigma_{\nu DM}=0$. Thus, one could expect bounds on $\sigma_0$ coming from this model to be particularly strong. Nevertheless, for a fixed $m_{DM}$, the constraints on $\sigma_0$ from IceCube, Xue~\textit{et al.}~\cite{Xue_2021} and Yang~\textit{et al.}~\cite{Yang:2024bsf} are usually within the ballpark of those from Wang~\textit{et al.}~\cite{Wang_2022}, and can be even stronger, regardless of them having a large number of events in the $\sigma_{\nu DM}=0$ scenario. This suggests that the number of events when $\sigma_{\nu DM}=0$ is not necessarily the most important factor in the placement of the limits, but rather the interplay between the spectrum of the flux and the scaling $n$ of $\sigma_{\nu DM}$, as shown when commenting the limits on $\mu$.

Another interesting fact we have shown is the uncertainty on the bound on $\sigma_0$ depends crucially on the column density, and thus on the assumed value of $R_{\textrm{em}}$. Moreover, models with a large $R_{\textrm{em}}$ have a very low sensitivity to particularities in the spike profile, and thus are likely to have low uncertainties. 

\subsection{Comparison with previous limits} \label{sec:comp}

\begin{figure}[tp]
\centering
\begin{subfigure}[b]{0.49\linewidth}
	\includegraphics[width=1\textwidth]{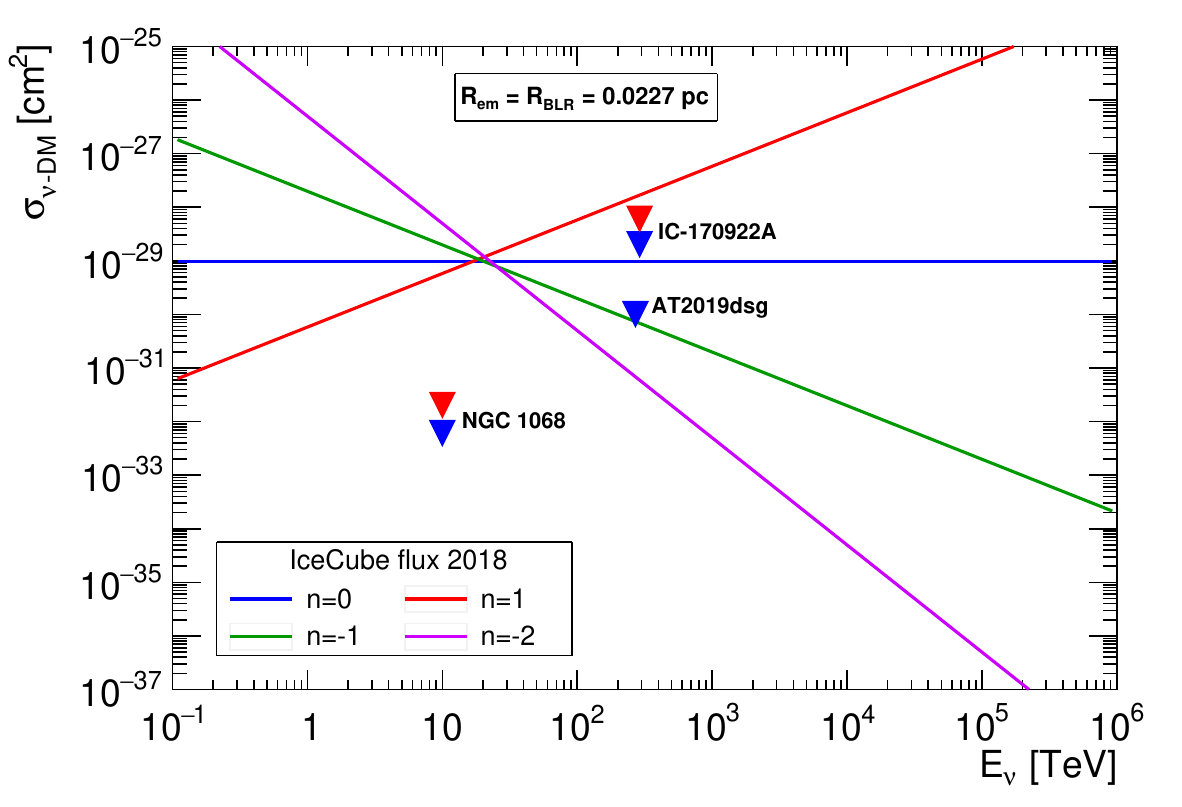}
	\caption{}
	\label{fig:xsec_IC_new}
\end{subfigure}
\begin{subfigure}[b]{0.49\linewidth}
	\includegraphics[width=1\textwidth]{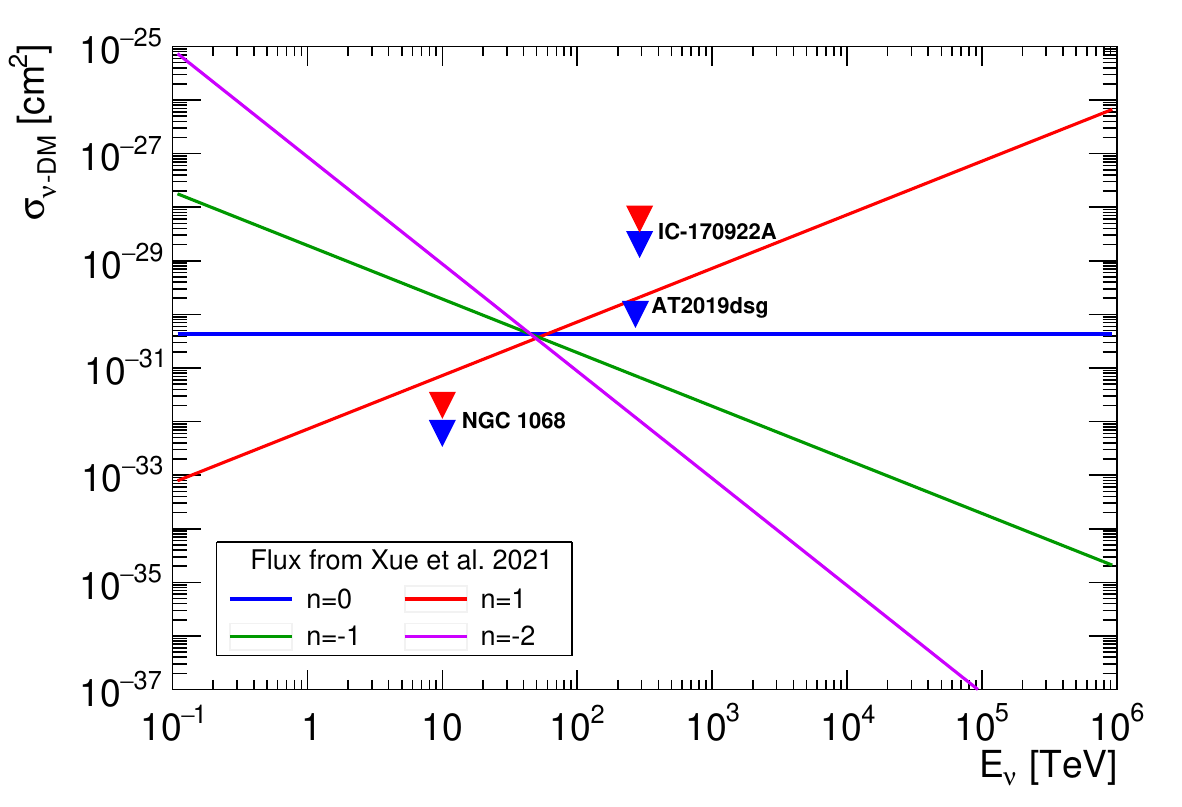}
	\caption{}
	\label{fig:xsec_xue}
\end{subfigure}
\begin{subfigure}[b]{0.49\linewidth}
	\includegraphics[width=1\textwidth]{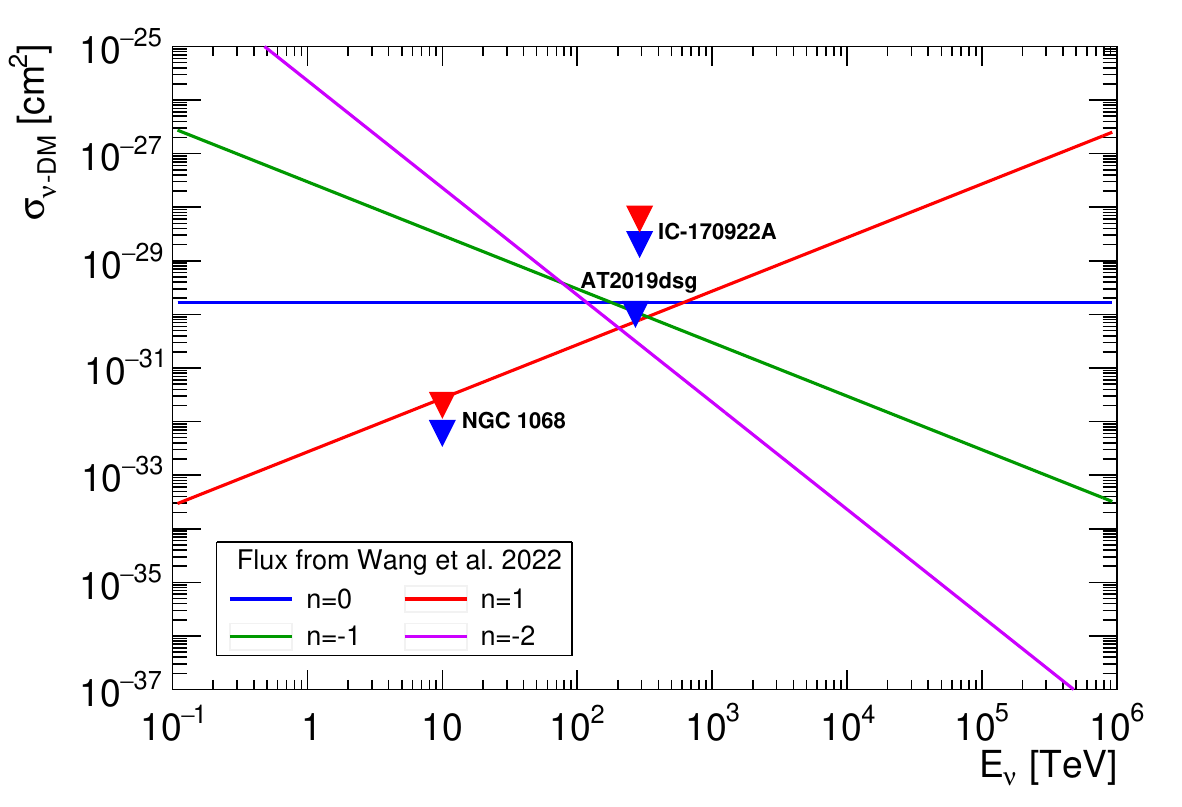}
	\caption{}
	\label{fig:xsec_wang}
\end{subfigure}
\begin{subfigure}[b]{0.49\linewidth}
	\includegraphics[width=1\textwidth]{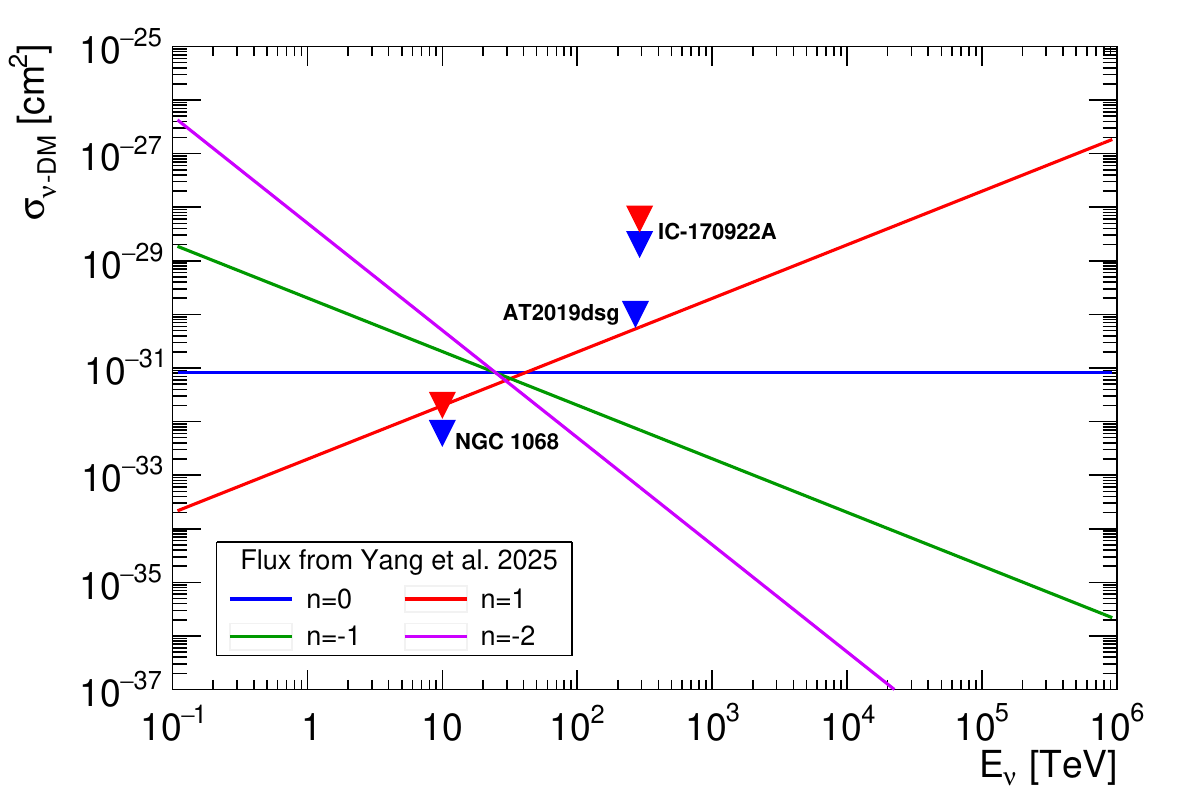}
	\caption{}
	\label{fig:xsec_yang}
\end{subfigure}
\caption{Bounds on $\sigma_{\nu DM}$ in function of neutrino energy $E_{\nu}$  obtained using the neutrino flux given in a) IceCube~\cite{Aartsen_2018b}, b) Xue~\textit{et al.}~\cite{Xue_2021}, c) Wang~\textit{et al.}~\cite{Wang_2022} and d) Yang~\textit{et al.}~\cite{Yang:2024bsf}; with their respective $R_{em}$. The dark matter mass is assumed to be $m_{DM} = 1$ GeV.}
\label{fig:Comparison1}
\end{figure}

As commented in the Introduction, neutrino - DM interactions have previously been examined in similar contexts, that is, astrophysical emission of neutrinos from SMBH, and their attenuation by the DM spike. The most restrictive constraints of this kind come from IceCube neutrino detections associated with NGC 1068 (an AGN, $E_{\nu} = 1.5-15$~TeV~\cite{Cline_2023}), the AT2019dsg event (from a TDE, $E_{\nu} \sim 270$~TeV~\cite{Fujiwara:2023lsv}), and the IC-170922A event (from blazar TXS~0506+056, $E_{\nu} \sim 290$~TeV~\cite{Cline_2022, Ferrer_2023}). It is thus desirable to compare these bounds with those derived in this work.

To this end, in Figure~\ref{fig:Comparison1} we presents our limits on the cross-section $\sigma_{\nu DM}$ as a function of the neutrino energy $E_{\nu}$, for $m_{DM} = 1$~GeV, obtained using the fluxes from IceCube~\cite{Aartsen_2018b}, Xue~\textit{et al.}~\cite{Xue_2021}, Wang~\textit{et al.}~\cite{Wang_2022} and Yang~\textit{et al.}~\cite{Yang:2024bsf}. These constraints are the strongest we have derived for each model, assuming the BM1 benchmark, that is $\gamma_{\textrm{sp}}=7/3$ and no DM self-annihilation. On each panel of the Figure, we present scenarios with $n=1,0,-1,-2$ in red, blue, green and purple, respectively.

The Figure also includes the aforementioned previous limits obtained from high-energy neutrinos. We begin our comparison with the limit from AT2019dsg~\cite{Fujiwara:2023lsv}. Being based on a single 270~TeV neutrino, and derived for $n=0$, it is presented on every panel as a blue triangle at the corresponding $E_\nu$. By comparing this with our blue lines, we find the bound from AT2019dsg to be stronger than those obtained at this value of energy from the IceCube and Wang~\textit{et al.}~\cite{Wang_2022} models, but weaker than Xue~\textit{et al.}~\cite{Xue_2021} and Yang~\textit{et al.}~\cite{Yang:2024bsf}. In all cases excepting IceCube, the difference is by less than an order of magnitude.

We now turn to the IC-170922A event, which again is based on a single $290$~TeV neutrino. In this case, the bound for $n=1$ is taken from~\cite{Cline_2022}, while that for $n=0$ is extracted from~\cite{Ferrer_2023}. Both are represented on each panel as red and blue triangles, respectively. We see that for the $n=0$ case, the limit is weaker than that from AT2019dsg~\cite{Fujiwara:2023lsv} by more than an order of magnitude, being also weaker than the bounds from all of our models. The constraint for $n=1$ case is weaker than $n=0$, but when compared with our limits, we see that for the corresponding energy it can be stronger than the bound derived from the IceCube flux, while being weaker than the rest.

We finish this part of the discussion by comparing the constraints from active galaxy NGC 1068. This time, we have a continuous emission of neutrinos, with IceCube observing around 80 events with energies between 1 - 15~TeV. Constraints for $n=1$ and $n=0$ come from~\cite{Cline_2023}, presented as usual as red and blue triangles in all panels, respectively, placed at the reference energy $E_0=10$~TeV used in~\cite{Cline_2023}. Here, we see that both cases place stronger constraints than the IceCube flux and the three models considered. For $n=1$, the models putting the closest constraint at the reference energy are those by Wang~\textit{et al.}~\cite{Wang_2022} and Yang~\textit{et al.}~\cite{Yang:2024bsf}, while for $n=0$ we have Yang~\textit{et al.}~\cite{Yang:2024bsf} slightly over one order of magnitude above it.

Interestingly, no other work has considered the $n=-1$ and $n=-2$ cases, which we consider particularly important. As we have mentioned before, most fluxes rely their prediction of $N_{\textrm{pred}}$ on the lowest energy part of the spectrum, which is very strongly affected by interactions going like $(E_\nu/E_0)^{-1}$ or $(E_\nu/E_0)^{-2}$.

Beyond astrophysical sources, a model-independent signal of interactions between DM and neutrinos is their effect on the CMB angular power spectra and the late-time matter power spectrum. 
For a constant cross section, Ref.~\cite{Mosbech:2020ahp} derives a limit of $\sigma_{\nu DM} < 2.2 \times 10^{-30}\,{\textrm{cm}}^2$~($m_{DM}/$GeV) using data from the CMB, baryon acoustic oscillations and gravitational lensing.
In addition, the study carried out in~\cite{Wilkinson:2014ksa} recognized the Lyman-$\alpha$ forest as a good probe of neutrino-DM interactions, setting a constraint on the cross-section of $\mathcal O(10^{-33})\,{\textrm{cm}}^2$, assuming it to be constant in energy and taking $m_{DM}=1$~GeV. However, a refined analysis in~\cite{Hooper:2021rjc} found a preference for non-zero DM-neutrino interactions\footnote{Further support for a possible nonzero interaction was found in an analysis based on CMB data from the Atacama Cosmology Telescope~\cite{Brax:2023rrf, Brax:2023tvn}, consistent with the Lyman-$\alpha$ result in~\cite{Hooper:2021rjc}.} so, to be conservative, we take this bound equal to their lower limit at 1$\sigma$, that is, $\sigma_{\nu DM} < 3.6 \times 10^{-32}\,{\textrm{cm}}^2$~($m_{DM}/$GeV).
A recent analysis using the observational data of Milky-Way satellite galaxies from the Dark Energy Survey (DES) and PanSTARRS1 obtain a bound of $\sigma_{\nu DM} <  10^{-32}\,{\textrm{cm}}^2$~($m_{DM}/$GeV) \cite{Akita:2023yga}.
Even stronger bounds have been obtained from 21~cm cosmology, reaching $\sigma_{\nu DM} < 4.4 \times 10^{-33}\,{\textrm{cm}}^2$~($m_{DM}/$GeV)~\cite{ Dey:2022ini}.

\begin{figure}
\centering
    \includegraphics[width=0.7\textwidth]{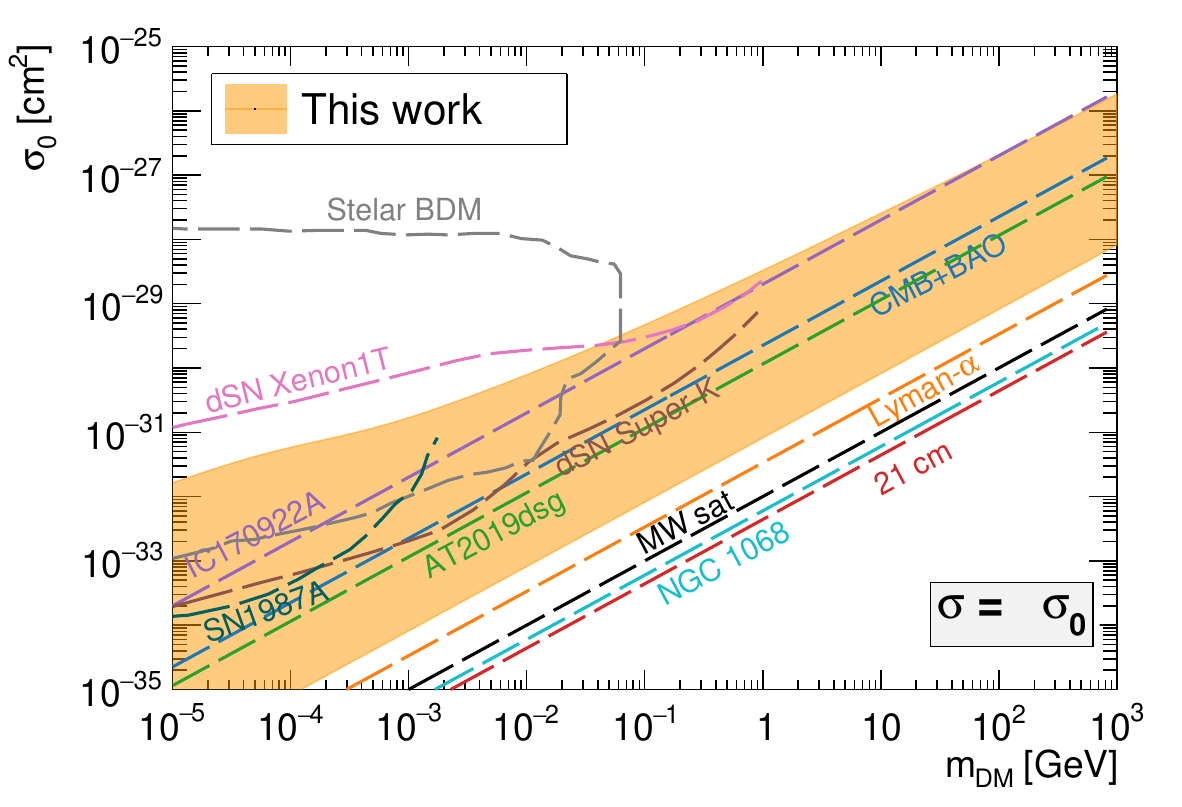}
\caption{\label{fig:xsec_cotas} Range of 90\% C.L. bounds on DM-neutrino cross section obtained considering energy-independent cross section, with previous constrains for comparison: (blue) CMB, baryon acoustic oscillations and lensing~\cite{Mosbech:2020ahp}; (orange) Lyman-$\alpha$ preferred model~\cite{Hooper:2021rjc}; (red) 21 cm cosmology \cite{ Dey:2022ini}; (black) Milky-Way satellite galaxies \cite{Akita:2023yga}; boosted dark matter searches with (pink, brown) diffuse supernova neutrinos~\cite{Ghosh:2021vkt}; (gray) stellar neutrinos~\cite{Jho:2021rmn}; (teal) supernova SN1987A~\cite{Lin:2022dbl}; (purple) bound from IC-170922A~\cite{Cline_2022, Ferrer_2023}; (green) tidal disrupted event AT2019dsg~\cite{Fujiwara:2023lsv}; (cyan) active galaxy NGC 1068~\cite{Cline_2023}.
}
\end{figure}

In addition to cosmological constraints, neutrino-DM interactions can be probed in direct detection experiments under the assumption that DM interacts with nucleons or leptons. High-energy neutrinos from stars~\cite{Jho:2021rmn}, diffuse supernovae~\cite{Ghosh:2021vkt}, or the supernova SN1987A~\cite{Lin:2022dbl} could boost DM particles, leading to an increased energy deposition from light DM candidates. Other methods to probe DM-neutrino interactions involve studying the attenuation of neutrino fluxes from supernovae, the Galactic Center, active galaxies, or tidal disruption events (TDEs).
Figure \ref{fig:xsec_cotas}, displays the most stringent limits for energy-independent cross-sections and compares them with the range of constrains derived in this work.
We can see that the strongest limits come from cosmology and from the AGN NGC 1068.

\section{Conclusions} \label{sec:concl}

With the true nature of dark matter remaining ever more elusive, it is important to gather data from all possible sources of interactions. In this work, we have used the neutrino flux observed during the 2014-2015 neutrino outburst of TXS 0506+056 to constrain the neutrino-DM cross-section. The logic behind the bound is that of preventing attenuation, which means that, given a particular neutrino flux coming from a blazar surrounded by a DM profile, one can place constraints on the neutrino-DM cross-section by demanding their interactions not to diminish the flux below a specific threshold.

An noteworthy difficulty found by most blazar models is to generate the high-energy neutrino flux while remaining consistent with X-ray and gamma-ray constraints. For our results, we took into account fluxes from three very different models giving a relatively large number of events at IceCube. We also considered the flux originally used by IceCube to fit their observed events. Then, by solving the cascade equation, we quantified the attenuation caused by both the DM spike and the DM halo, considering various energy-dependent cross-sections motivated by simplified models.

To account for uncertainties in the DM spike parameters, we present our results as a range of bounds, obtained by comparing the most and least restrictive constraints. We find that the limits on the reference cross-section $\sigma_0$ are highly sensitive to the assumed neutrino flux model, the DM particle mass $m_{DM}$, the properties of the DM spike (such as its slope $\gamma_{\textrm{sp}}$ and the DM self-annihilation cross-section $\langle\sigma v \rangle_{\textrm{ann}}$), and the energy scaling $n$ of the neutrino-DM cross-section. 

The main results of this work can be found in Figure~\ref{fig:regions}, which shows the uncertainty bands as a function of $m_{DM}$ for each blazar model and for each assumed $n$ for the cross-section. We find that for a fixed $m_{DM}$, the uncertainty can span between $\sim2$ and $\sim5$ orders of magnitude, with the strongest bounds being around $\mathcal O(10^{-37})\,{\textrm{cm}}^2$ ($\mathcal O(10^{-29})\,{\textrm{cm}}^2$) for $m_{DM}=10$~keV (1~TeV), for the model of Yang~\textit{et al.}~\cite{Yang:2024bsf} and $n=-2$.

Our results are compared with other constraints in Figs.~\ref{fig:Comparison1} and~\ref{fig:xsec_cotas}. Although our limits are weaker than those derived from NGC 1068, they are comparable to, and in some cases more restrictive than, those obtained from TDEs and the IC-170922A event. Notably, no previous study has explored cross-sections with $n=-1$ or $n=-2$, cases that we introduce in this work, motivated by models involving light scalar mediators.

Looking ahead, future data from Baikal-GVD and KM3NeT, together with continued IceCube observations, will enable the identification of more high-energy neutrino sources. This will further improve our understanding of neutrinos and their interactions with dark matter.

\section*{Acknowledgments}

We gratefully acknowledge financial support from the Dirección de Fomento de la Investigación at Pontificia Universidad Católica del Perú through Grant No. DFI-PUCP PI0758, as well as from the Vicerrectorado de Investigación at Pontificia Universidad Católica del Perú via the Estancias Posdoctorales en la PUCP 2023 program.

GZ gratefully acknowledges funding from the European Union Horizon Europe research and innovation programme under the Marie Sklodowska-Curie Staff Exchange grant agreement No. 101086085 – ASYMMETRY, which supported his participation in the Invisibles Workshop 2024 in Bologna, providing a valuable opportunity for discussions. GZ also acknowledges financial support from the Dirección de Fomento de la Investigación at Pontificia Universidad Católica del Perú through Grant No. PI1144, which enabled his participation in SILAFAE 2024 in Mexico City, fostering further discussions.

 \bibliographystyle{JHEP}
 \bibliography{biblio.bib}

\end{document}